\documentclass{article}

\usepackage{arxiv}

\usepackage[utf8]{inputenc} 
\usepackage[T1]{fontenc}    
\usepackage[hidelinks]{hyperref}       
\usepackage{url}            
\usepackage{booktabs}       
\usepackage{amsfonts}       
\usepackage{nicefrac}       
\usepackage{microtype}      
\usepackage{lipsum}		
\usepackage{graphicx}
\usepackage{doi}

\usepackage{float}

\usepackage{{ulem}}
\usepackage{siunitx}
\usepackage{textcomp, gensymb}
\usepackage[xindy,style=long,toc=true,nonumberlist,acronym,acronymlists={abbreviations},hyperfirst=true]{glossaries}
\usepackage{lscape}

\usepackage{multirow}
\usepackage[title]{appendix}

\newacronym{DC}{DefChar}{\textit{defect characteristic}}
\newacronym{IR}{$\mathrm{ImR}$}{\textit{image retrieval}}
\newacronym{CBIR}{CBIR}{\textit{content-based image retrieval}}
\newacronym{CNN}{CNN}{\textit{convolutional neural network}}
\newacronym{mAP}{$\mathrm{mAP}$}{\textit{mean average precision}}
\newacronym{LBP}{LBP}{\textit{local binary pattern}}
\newacronym{AI}{AI}{\textit{artificial intelligence}}
\newacronym{SIFT}{SIFT}{\textit{scale invariant feature transform}}
\newacronym{SAM}{SAM}{\textit{spectral angle mapper}}
\newacronym{UQI}{UIQ}{\textit{universal image quality index}}
\newacronym{MSE}{MSE}{\textit{mean squared error}}
\newacronym{SSIM}{SSIM}{\textit{structural similarity index}}
\newacronym{AP}{$\mathrm{AP}$}{\textit{average precision}}
\newacronym{CT}{CT}{computerised tomography}

\newcommand{\etal}{\text{\textit{et al.\@ }}}
\newcommand{\etals}{\text{\textit{et al.\@}'s }}

\title{Efficient Retrieval of Images with Irregular Patterns using Morphological Image Analysis: Applications to Industrial and Healthcare datasets}

\author{Jiajun Zhang\thanks{Corresponding author},\hspace{1mm} Georgina Cosma$^{*}$ \\
	Department of Computer Science\\
        School of Science\\
	Loughborough University\\
	Loughborough, LE11 3TU, UK \\
	\texttt{j.zhang8@lboro.ac.uk}$^{*}$ \\
        \texttt{g.cosma@lboro.ac.uk}$^{*}$ \\
        \And
        Sarah Bugby \\
	Department of Physics\\
        School of Science\\
	Loughborough University\\
	Loughborough, LE11 3TU, UK \\
	\texttt{s.bugby@lboro.ac.uk} \\
        \And
        Jason Watkins \\
	Railston \& Co. Ltd.\@\\
        Nottingham, NG7 2TU\\
	\texttt{jason@railstons.com} \\
}

\date{}

\renewcommand{\shorttitle}{}

\hypersetup{
pdftitle={Efficient Retrieval of Images with Irregular Patterns using Morphological Image Analysis: Applications to Industrial and Healthcare datasets},
pdfsubject={},
pdfauthor={Jiajun Zhang, Georgina Cosma, Sarah Bugby and Jason Watkins},
pdfkeywords={image retrieval, morphological defect characteristics, irregular pattern analysis},
}

\begin{document}
\maketitle

\begin{abstract}
Image retrieval is the process of searching and retrieving images from a database based on their visual content and features. Recently, much attention has been directed towards the retrieval of irregular patterns within industrial or medical images by extracting features from the images, such as deep features, colour-based features, shape-based features and local features. This has applications across a spectrum of industries, including fault inspection, disease diagnosis, and maintenance prediction. This paper proposes an image retrieval framework to search for images containing similar irregular patterns by extracting a set of morphological features (DefChars) from images; the datasets employed in this paper contain wind turbine blade images with defects, chest computerised tomography scans with COVID-19 infection, heatsink images with defects, and lake ice images. The proposed framework was evaluated with different feature extraction methods (DefChars, resized raw image, local binary pattern, and scale-invariant feature transforms) and distance metrics to determine the most efficient parameters in terms of retrieval performance across datasets. The retrieval results show that the proposed framework using the DefChars and the Manhattan distance metric achieves a mean average precision of \SI{80}{\percent} and a low standard deviation of 0.09 across classes of irregular patterns, outperforming alternative feature-metric combinations across all datasets. Furthermore, the low standard deviation between each class highlights DefChars' capability for a reliable image retrieval task, even in the presence of class imbalances or small-sized datasets.
\end{abstract}


\section{Introduction}
\label{sec:intro}

\Gls{IR} is the task of searching and analysing images; some applications include face recognition, image search engines, image metadata annotation, object classification, and more. Recently, \gls{IR} has been applied to retrieve similar images based on their irregular patterns for fault inspection and disease diagnosis purposes. Irregular pattern analysis can detect industrial defects \cite{01_wafer, 02_fabric, 04_bridge}, chest infections in medical scans \cite{05_chestxray, 20_ct, 21_lung}, and ice or snow on lakes \cite{ex1_ice,ex2_ice}, serving industry, healthcare, and environmental monitoring. An accurate \gls{IR} system can aid experts (e.g.\ manufacturing engineers, doctors, quality inspectors, security officers, etc.\@) during decision-making.

Many research studies have explored the retrieval of images containing irregular patterns in industrial or medical datasets using different features and similarity metrics. Image-based similarity metrics (e.g.\ \gls{MSE}, \gls{UQI} \cite{11_uqi}, \gls{SAM} \cite{12_sam}, etc.\@) \cite{29_mse_sim, 13_ssim, 30_ct_sim, 31_sim}, which compare the similarity between image data, provide a simple and intuitive means of comparing two images in \gls{IR} tasks. However, the similarity values computed from these metrics are sensitive to image noise and quality. Feature extraction methods can extract the hidden features of the irregular patterns within images and improve retrieval performance. These methods extract \gls{LBP} features \cite{22_wlbp_steeldefect,32_lbp_fea,33_lbp_fea,34_lbp_fea}, \gls{SIFT} features \cite{35_sift_fea,36_sift_fea, 37_sift_fea}, as well as color and shape features \cite{16_simfea1, 27_fesim1} to conduct retrieval of images with irregular patterns. Distance-based similarity metrics (e.g.\ Manhattan, Jaccard, Euclidean, Cosine, etc.\@) can be utilised to compute similarity values between two sets of features extracted from images with irregular patterns. Zhang \etal \cite{08_morfea} proposed a set of morphological features, known as \glspl{DC}, to characterise images with irregular patterns in terms of colour, shape, and meta aspects. Zhang \etal \cite{08_morfea} successfully utilised the \glspl{DC} to reason the outputs from an artificial intelligence-based defect detection and classification model. However, using \glspl{DC} as features in an \gls{IR} task is still a question waiting to be explored.

This paper proposes an \gls{IR} framework that can extract \glspl{DC} from images containing irregular patterns and retrieve images with similar irregular patterns by comparing their \glspl{DC} vectors using a feature-based similarity metric. Four datasets are employed in this study: wind turbine blade images with defects, chest \gls{CT} scans with COVID-19 infection, heatsinks images with defects, and lake images with ice. The proposed framework is evaluated with different feature-metric combinations, such as \glspl{DC} vectors and feature-based similarity metrics, resized raw images and image-based similarity metrics (\gls{MSE}, \gls{SAM}, \gls{UQI}), \gls{LBP} and feature-based similarity metrics, \gls{SIFT} and the Euclidean metric. The retrieval results demonstrate that using the combination of \glspl{DC} and the Manhattan metric within the proposed framework consistently achieves the highest \gls{mAP} (average 0.80) and also maintains the lowest standard deviation (average 0.09) across classes of irregular patterns and fast retrieval time (average 0.14 seconds per query) across all datasets. Additionally, the retrieval results also indicate that using \glspl{DC} within the proposed framework can relatively have high and balanced retrieval accuracy across classes despite dataset imbalances or small-sized datasets. The proposed \gls{IR} framework could be expanded to various industrial tasks in the future, including irregular pattern identification, classification, deterioration monitoring, repairing predictions and more.

There are six sections in this paper. Section \ref{sec:rw} interprets related works concerning similarity metrics and feature extraction methods that can be employed in the \gls{IR} task. Section \ref{sec:arch} presents the proposed \gls{IR} framework for retrieving irregular patterns in industrial or medical datasets. Section \ref{sec:meth} outlines the datasets, relevant feature extraction methods, similarity metrics, and the methodology used in this research, providing insight into the experimental setup. Section \ref{sec:res} evaluates and discusses the retrieval performance and execution time of various features and similarity metrics explored in the paper. Section \ref{sec:conc} summarises the key findings and conclusions drawn from the research conducted in this paper; moreover, some future works are briefly pointed out.

\section{Related Works}
\label{sec:rw}

\subsection{Feature Extraction and Relevant Similarity Metrics for Retrieving Images}

Recently, an increasing number of researchers have explored how effective feature extraction methods can enhance the performance of \gls{IR}. In 2019, Latif \etal \cite{07_fereview} provided a comprehensive review of successful feature extraction methods used in \gls{CBIR} tasks. There are six major types of features that can be extracted from these methods, including colour-based features, shape-based features, texture-based features, spatial features, fusion features, and local features. \textit{Colour-based features} \cite{color1,color2,color3,color4,color5,color6,color7,color8} offer fundamental visual information that is similar to human vision, and they are relatively robust against image transformations. \textit{Texture-based features} \cite{texture1,texture2,texture3,texture4,texture5,texture6,texture7} capture repeating patterns of local variance in image intensity; these features often hold more semantic meaning than colour-based features, though they can be susceptible to image noise. \textit{Shape-based} features decode an object's geometrical forms into machine-readable values; Latif \etal \cite{07_fereview} summarised that shape-based features can encompass contour, vertex angles, edges, polygons, spatial interrelation, moments, scale space, and shape transformation. \textit{Spatial features} \cite{spitial1,spitial2,spitial3,spitial4,spitial5,spitial6,spitial7} convey the location information of objects within the image space. \textit{Fusion features} \cite{fusion1,fusion2,fusion3,fusion4,fusion5} combine basic features to form high-dimensional concatenated features; often, principal component analysis is applied to reduce dimensions. \textit{Local features} \cite{local1,local2,local3,local4,local5,local6,local7,local8} represent distinct structures and patches in an image, providing fine-grained details for \gls{IR} tasks. 

Seetharaman and Sathiamoorthy \cite{27_fesim1} applied the Manhattan similarity metric along with colour-based and shape-based features to complete a medical \gls{IR} task. The results demonstrated that their method achieved the highest average retrieval rate of \SI{84.47}{\percent} and a speed of 2.29 seconds. Petal et al.\@ \cite{16_simfea1} extracted both colour-based and texture-based features from images and applied them to a \gls{CBIR} task using various distance measures (e.g., Euclidean, Cosine, Jaccard, Manhattan, etc.\@). Their approach achieved an impressive accuracy of \SI{87.2}{\percent} in retrieving similar images. In 2022, Shamna et al.\@ \cite{28_fesim2} employed the bag of visual words model as spatial features for retrieving medical images. Their method excelled in handling grayscale datasets, achieving a \gls{mAP} of \SI{69.70}{\percent}. However, its performance on coloured datasets was less satisfactory.

\subsection{Recent works to retrieve industrial and medical images with irregular patterns}

In 2021, Boudani et al.\@ \cite{22_wlbp_steeldefect} employed wavelet-based \gls{LBP} feature with the chi-square similarity metric to identify images containing surface defects on hot-rolled steel strips, achieving an $\mathrm{mAP}@10$ score of 0.93; however, performance was not stable between classes. Mo et al.\@ \cite{24_fabric01} proposed a concentrated hashing method with neighbourhood embedding, utilising a \gls{CNN} to extract hashing features, for retrieving fabric and textile datasets in industrial applications. Their method outperformed other methods in four fabric datasets with an average \gls{mAP} of over \SI{90}{\percent}, but the precision sharply dropped by \SI{35}{\percent} when retrieving more than 8 images. In 2022, Deep et al.\@ \cite{25_med} introduced a texture descriptor based on the concept of \gls{LBP} for conducting \gls{IR} tasks on three biomedical datasets. The results showed that their proposed methods reached an \gls{AP} rate of \SI{91.5}{\percent}. However, the \gls{AP} rate of one of the datasets was much lower than other datasets (i.e.\ \SI{93}{\percent}, \SI{90}{\percent}, \SI{46}{\percent}). Furthermore, their proposed descriptor needs longer retrieval times because the descriptor had a large vector size ($3 \times 4 \times 256$). Maintaining a consistent retrieval performance using the same feature and similarity metric for different datasets is challenging in the \gls{IR} domain. Boudani et al.\@ \cite{22_wlbp_steeldefect} and Deep et al.\@ \cite{25_med} both applied the \gls{LBP}-based method for \gls{IR} tasks, but their \glspl{mAP} differed across their respective datasets.

Zhang \etal \cite{08_morfea} introduced a set of 38 morphological features named \glspl{DC}, which capture attributes related to defects in terms of colour, shape, and meta characteristics. This feature set explains defect attributes through visualisations, enhancing understanding for individuals with human visual knowledge. Furthermore, Zhang \etal \cite{08_morfea} employed \glspl{DC} in an \gls{AI} reasoning task, showcasing its capability in data explanation and reasoning. Despite this advancement, there remains a lack of research focused on evaluating retrieval performance using the \glspl{DC}.


\section{Proposed Image Retrieval Framework}
\label{sec:arch}

\begin{figure}[!ht]
\includegraphics[width=1\textwidth]{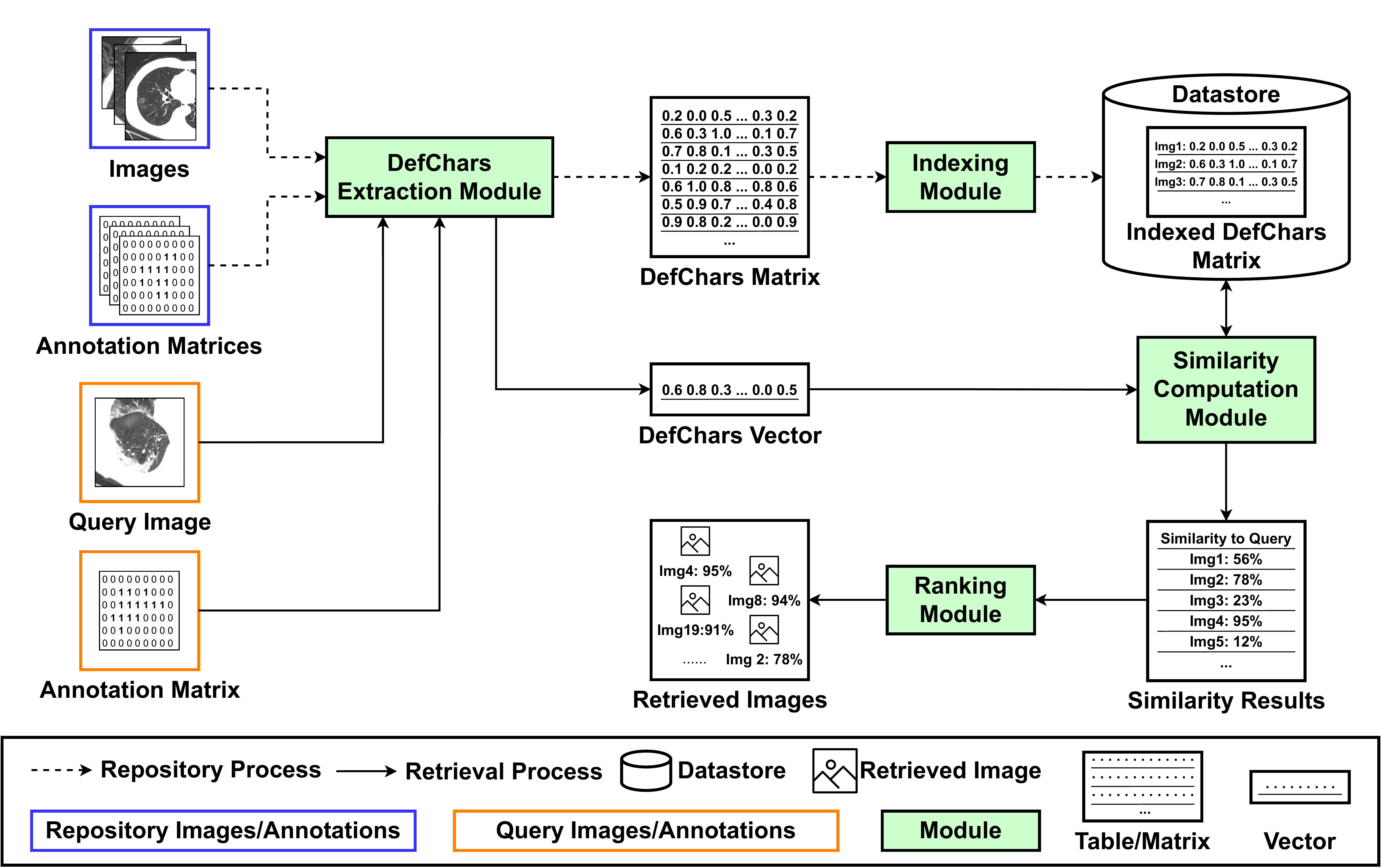}

\caption{Proposed Image Retrieval Framework.}
\label{fig:arch}
\end{figure}

This section introduces a proposed \gls{IR} framework, as shown in Figure \ref{fig:arch}, retrieving images with similar irregular patterns in industrial or medical datasets through the use of \glspl{DC}. The code for our proposed \gls{IR} framework is provided in a GitHub repository\footnote{https://github.com/edgetrier/ImgRetreival-DefChars}. It consists of two main processes: the repository process and the retrieval process. The repository process employs two key modules, the \glspl{DC} Extraction module and the Indexing module, to construct a datastore. This datastore contains a \glspl{DC} matrix extracted from annotated images that include irregular patterns. The retrieval process, on the other hand, relies on three modules: the \glspl{DC} Extraction module, the Similarity Computation module, and the Ranking module. These modules are used to search for the images having similar irregular patterns within the datastore by comparing the extracted \glspl{DC} vectors.

\subsection{Repository Process}
\label{sec:reppro}

\textbf{\glspl{DC} Extraction module:} This module serves as a feature extraction component responsible for generating a \glspl{DC} matrix that extracts the colour-based, shape-based, and meta-based features of irregular patterns within images. It is important to note that this module can be replaced by raw image data or other feature extraction methods, such as \gls{LBP} or \gls{SIFT}. The input to this module consists of a set of \textit{images} and corresponding \textit{annotation matrices}. Each image in the input set is required to contain a single irregular pattern. Additionally, the corresponding annotation is represented as a mask-based matrix, outlining the irregular pattern's region within the image. This matrix matches the size of the input image, with each value indicating whether the corresponding pixel in the input image falls inside or outside the irregular pattern's region. To prepare the annotation matrices, image annotation tools like \textit{VIA} \cite{via}, \textit{Labelme} \cite{labelme}, or the \textit{drawContours} function from the \textit{OpenCV} package \cite{opencv} can be employed if the dataset lacks annotation data. Subsequently, the module computes a \glspl{DC} vector (size: $38 \times 1$) for each image to represent the \glspl{DC} of the irregular pattern. This is achieved by analysing the pixel values within both the irregular pattern and background regions, as detailed in Table \ref{tab:defchar}. The values in these vectors are then normalised to a range between 0 and 1. Finally, these \glspl{DC} vectors are aggregated into a \textit{\glspl{DC} matrix}, which serves as the module's output.

\textbf{Indexing module:} This module associates each input image with its respective \glspl{DC} vector during the repository process. It takes the extracted \textit{\glspl{DC} matrix} as input and assigns a unique index to each vector within the \glspl{DC} matrix. The index within a \glspl{DC} vector can assist users in locating the corresponding input image during the retrieval process. The module outputs an indexed \glspl{DC} matrix and stores it in a \textit{datastore}. Additionally, this module can extend its indexing functionality when adding new images to the datastore.

\subsection{Retrieval Process}
\label{sec:retpro}

\textbf{\glspl{DC} Extraction module:} This module is the same as the one described in Section \ref{sec:reppro}. However, its function is to extract a single \glspl{DC} vector from an annotated query image. The input for this module consists of an \textit{image} containing a single irregular pattern and an \textit{annotation matrix} showing the region of the irregular pattern within the image. The module generates a single \textit{\glspl{DC} vector} that represents the features of the irregular pattern within the query image.

\textbf{Similarity Computation module:} This module compares the \glspl{DC} vector extracted from the query image to each \glspl{DC} vector in the datastore using a feature-based similarity metric. In this paper, the Manhattan metric was employed as the feature-based similarity metric, although it can be substituted with any distance-based metric (e.g.\ Cosine, Jaccard, Euclidean, etc.\@). The input to this module comprises the \textit{datastore} (\glspl{DC} matrix) and the \textit{\glspl{DC} vector} extracted from the query image. Subsequently, the module computes a set of similarity values using the selected metric to illustrate the similarity between each \glspl{DC} vector in the datastore and the \glspl{DC} vector extracted from the query image. Finally, the set of similarity values is recorded and outputted in a \textit{Similarity Results} table.

\textbf{Ranking module:} This module is the last step of the retrieval process, which ranks the retrieved results and presents the retrieved images. The module's input is the \textit{Similarity Results} table generated from the similarity computation module. Next, the Similarity Results table is ranked in order of the computed similarity values, and the module outputs the respective \textit{images} according to their index.

\section{Experiment Methodology}
\label{sec:meth}

\subsection{Datasets}
\label{sec:datasets}

\begin{table}[!ht] 
\caption{Class Distribution of irregular patterns in Each Dataset; -- represents that there is no such type in the dataset.}
\label{tab:typedist}
\centering
\footnotesize
\begin{tabular}{lcccccc}
\toprule
\multirow{2}{*}{Dataset} & Number of & \multicolumn{4}{c}{Number of Irregular Patterns} & Total \\ \cmidrule{3-6}
 & Images  & Class 1 & Class 2 & Class 3 & Class 4 & Irregular Patterns \\\midrule
Wind Turbine Blade Defect & 191 & 89 & 73 & 118 & 24 & 304\\
Lake Ice \cite{19_lakeicedata} & 4017 & 606 & 1207 & 3237 & 315 & 5365 \\
Chest \gls{CT} \cite{17_chestCTdata} & 750 & 2317 & 1668 & 680 & -- & 4665 \\
Heatsink Defect \cite{18_heatsinkdata} & 1000 & 2160 & 4927 & -- & -- & 7007 \\
\bottomrule
\end{tabular}
\end{table}

\begin{figure}[!ht]
\centering
\includegraphics[width=\linewidth]{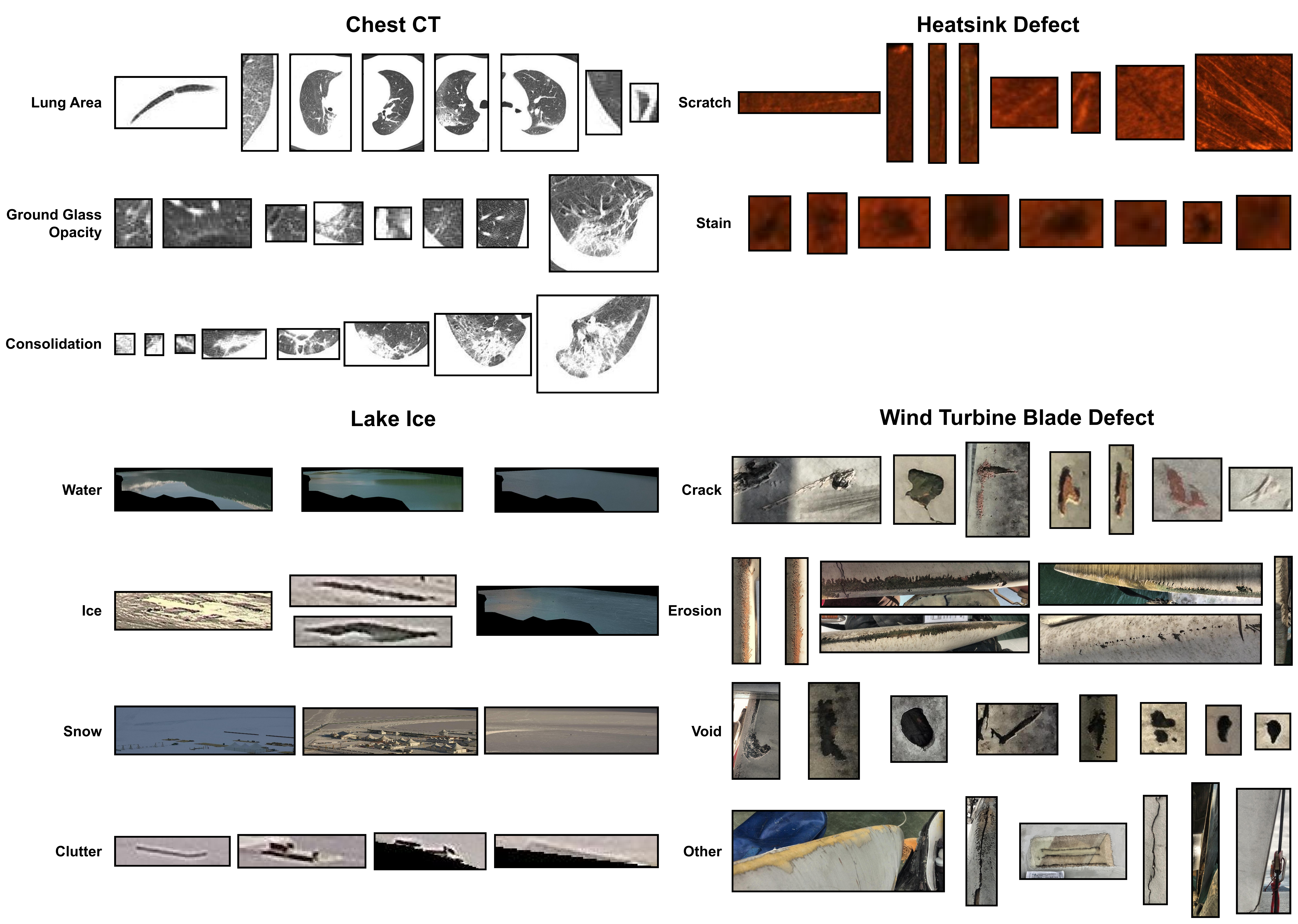}

\caption{Example images of irregular patterns in each class of every dataset.}
\label{fig:expimg}
\end{figure}  

The experiment applied the proposed \gls{IR} framework to four distinct datasets: wind turbine blade images with defects, chest \gls{CT} images with lung infections, heatsink images with defects, and lake images with ice. The wind turbine blade defect images were provided by the industrial partner \textit{Railston \& Co Ltd.}; the chest \gls{CT} images were sourced from Ter-Sarkisov's \cite{17_chestCTdata} study; the heatsink defect images were gathered from Yang \etals experiment \cite{18_heatsinkdata}; the lake ice images were collected from the work of Prabha \etal \cite{19_lakeicedata}. All of these datasets include mask annotations that identify the regions and classes of irregular patterns.

To ensure comprehensive irregular pattern retrieval within these datasets, each irregular pattern was cropped into an individual image based on the boundaries outlined in its respective mask annotation. The distribution of classes within each dataset is outlined in Table \ref{tab:typedist}. Moreover, Figure \ref{fig:expimg} showcases select cropped images for each class within each dataset.

\subsection{Feature Extraction Method for Image Retrieval}
\label{sec:fesm}

\begin{table}[p]
\caption{Proposed \glspl{DC}.} 
\centering
\footnotesize
\renewcommand{\arraystretch}{1.3}
\begin{tabular}{lcp{3.7in}}
\toprule
 \multicolumn{3}{c}{\textbf{Colour Information extracted and stored separately for the defect and background areas}} \\ \midrule
 \textbf{DefChar Name} & \textbf{Value Range} & \textbf{Description} \\ \midrule
 Average Hue & $\{0,1,\dotsc, 359\}$ & Average hue value \\ 
 Mode of Hue & $\{0,1,\dotsc, 359\}$ & Most frequent hue value \\ 
 Unique number of Hue values & $\{1,2,\dotsc, 360\}$ & Number of unique hue values \\
 Hue Range & $\{0,1,\dotsc, 180\}$ &  Difference of maximum and minimum hue value  \\
 Average Saturation & $\{0,1,\dotsc, 254\}$ & Average saturation value \\
 Mode of Saturation & $\{0,1,\dotsc, 254\}$ & Most frequent saturation value \\
 Unique number of Saturation & $\{1,2,\dotsc, 255\}$ & Number of unique saturation values \\ 
 Saturation Range & $\{0,1,\dotsc, 254\}$ & Difference of maximum and minimum saturation values \\
 Average Brightness & $\{0,1,\dotsc, 254\}$ & Average brightness value \\
 Mode of Brightness & $\{0,1,\dotsc, 254\}$ & Most frequent brightness value \\
 Unique number of Brightness & $\{1,2,\dotsc, 255\}$ & Unique brightness values \\
 Brightness Range & $\{0,1,\dotsc, 254\}$ & Difference of maximum and minimum brightness value \\\midrule
 \multicolumn{3}{c}{\textbf{Colour Complexity}} \\\midrule
 \textbf{DefChar Name} & \textbf{Value Range} & \textbf{Description} \\ \midrule
 Hue Difference & $[0,1]$ & Hue frequency distribution difference between the defect and background areas \\
 Saturation Difference & $[0,1]$ & Saturation frequency distribution difference between the defect and background areas \\
 Brightness Difference & $[0,1]$ & Brightness frequency distribution difference between the defect and background areas \\\midrule
 \multicolumn{3}{c}{\textbf{Shape Information}} \\ \midrule
 \textbf{DefChar Name} & \textbf{Value Range} & \textbf{Description} \\ \midrule
 Number of Edges & $\{3,4,\dotsc\}$ & Number of edges of the defect polygon areas \\
 Coverage & $[0,1]$ & Percentage of the defect polygon area covered by its bounding box \\
 Aspect Ratio & $[0,1]$ & Ratio between the width and height of defect bounding box \\
 Average Turning Angles & $\{1,2,\dotsc, 180\}$ & Average value of vertex angles of the defect polygon area \\
 Mode of Turning Angle & $\{1,2, \dotsc, 180\}$ & Value of vertex angles that appears the most often in the defect polygon \\ \midrule
 \multicolumn{3}{c}{\textbf{Shape Complexity}} \\ \midrule
 \textbf{DefChar Name} & \textbf{Value Range} & \textbf{Description} \\ \midrule
 Edge Ratio & $[0,1]$ & Average length ratio between two adjacent edges in the defect polygon area \\
 Followed Turns & $[0,1]$ & Proportion of two adjacent vertices which turn to the same direction in the defect polygon area \\
Small Turns & $[0,1]$ & Percentage of vertices which are smaller than 90$\degree$ in the defect polygon area \\
 Reversed Turns & $[0,1]$ & Proportion of two adjacent vertices which turn to a different direction in the defect polygon area \\\midrule
 \multicolumn{3}{c}{\textbf{Meta Information}} \\ \midrule
 \textbf{DefChar Name} & \textbf{Value Range} & \textbf{Description} \\ \midrule
 Defect Size & $\{1,2,\dotsc\}$ & Number of pixels in the defect polygon area \\
 Neighbour Distance & $\{0,1,2\}$ & Categorised distances to the nearest neighbour, 0$\rightarrow$Short ($\leq$\SI{100}{px}); 1$\rightarrow$Long; 2$\rightarrow$No Neighbour. \\\bottomrule
\end{tabular}
\label{tab:defchar}
\end{table}

\textbf{Defect Characteristics:} This feature extraction methodology necessitates images with associated mask-based annotations, which are essential for calculating the \gls{DC} values corresponding to each defect within the dataset. This is particularly important since a single image may encompass multiple defects. Table \ref{tab:defchar} provides a comprehensive list of the \gls{DC}, their respective value ranges, and descriptions. The outcome of Zhang \etal's method \cite{08_morfea} manifests as a matrix with dimensions $38 \times n$, where $n$ signifies the count of defects present within the dataset.

\textbf{Scale Invariant Feature Transform:} Lowe \cite{09_sift} introduced \gls{SIFT}, a local feature extraction method which can maintain an image's scale invariance. \gls{SIFT} identifies a set of keypoints and descriptors, capturing distinctive points within images. This set of keypoints and descriptors can subsequently be employed to compute similarity between images by comparing their keypoints and descriptors using the Euclidean distance metric. The \gls{SIFT} extraction method contains four key steps. The first step, scale-space extrema detection, utilises the Gaussian pyramid, images are progressively downsampled to identify a collection of potential keypoints. This is achieved by analysing the differences across each level of the Gaussian pyramid. The second step, orientation assignment, involves the elimination of low-contrast keypoints, enhancing the quality of the selected keypoints. The third step, keypoint descriptor computes dominant orientations for individual keypoints, ensuring invariance to image rotation. Lowe \cite{09_sift} recommended using Euclidean similarity metric (explained in Section \ref{sec:feasm}) to determine the similarity between the keypoints and descriptors of two images.

\textbf{Local Binary Pattern:} Ojala \etal \cite{10_lbp} introduced a texture descriptor feature extraction method, designed for \gls{IR} tasks. The \gls{LBP} method extracts texture descriptors by comparing the intensity value of each pixel in an image to the intensity values of its neighbouring pixels. The \gls{LBP} extraction method contains four sequential steps. The first step, neighbourhood definition, identifies the eight neighbouring pixels around each pixel within the image. The second step, binary comparison, calculates the binary intensity value of each neighbouring pixel, relative to the centre pixel. The third step, binary pattern generation, concatenates all the binary intensity values into a singular vector, following either a clockwise or counter-clockwise order. The last step, decimal Representation, converts the generated binary patterns into a decimal number, serving as a representation of the texture feature.

\subsection{Similarity Metrics for Image Retrieval}
\label{sec:simmea}

There are two categories of similarity metrics to conduct an \gls{IR} task: image-based metrics for raw image data and feature-based metrics for extracted feature data. Image-based metrics compute the similarity or dissimilarity between images by directly comparing the raw image data. \gls{MSE}, \gls{SAM} and \gls{UQI} are used in this experiment. On the other hand, feature-based metrics assess the similarity or dissimilarity between images by analysing the extracted features. Euclidean, Cosine, Jaccard and Manhattan distance metrics are utilised in this experiment. In the descriptions that follow, let $X$ be the query image; and let $Y_p$ be one of the retrieved images.

\subsubsection{Image-based Similarity Metrics}
\label{sec:imgsm}

Image similarity metrics play a crucial role in an \gls{IR} task by searching similar images within a database. Traditional image similarity metrics, such as \gls{MSE}, \gls{SAM} \cite{12_sam}, \gls{UQI} \cite{11_uqi}, and \gls{SSIM} \cite{13_ssim}, allow for a direct comparison of pixel value differences between two images using mathematical equations.

\textbf{Mean Square Error (MSE)} calculates the average squared difference in pixel values between two images. A higher \gls{MSE} value signifies a greater dissimilarity between the two images.
\begin{equation}
   \mathrm{MSE} = \frac{1}{HWC}\sum_{i=1}^{H}\sum_{j=1}^{W}\sum_{k=1}^{C}(x(i,j,k)-y(i,j,k))^2
\end{equation}
where $H$ represents the height of the image; $W$ represents the width of the image; $C$ represents the number of channels (colour components) in each pixel;  $x(i,j,k)$ represents the pixel value of the $i$th row, $j$th column, and $k$th channel in the query image $X$; $y(i,j,k)$ represents the pixel value of the $i$th row, $j$th column, and $k$th channel in the retrieving image $Y$.

\textbf{Spectral Angle Mapper (SAM)} calculates the angular disparity between two spectral signatures within a high-dimensional spectral space. A higher \gls{SAM} value signifies a greater dissimilarity between the two images.

\begin{equation}
   \mathrm{SAM} = \frac{1}{C}\sum_{k=1}^{C}cos^{-1}\left \{\frac{\sum_{i=1}^{H}\sum_{j=1}^{W}x(i,j,k) \cdot y(i,j,k)}{\sqrt{\sum_{i=1}^{H}\sum_{j=1}^{W}x(i,j,k)^2}\cdot 
 \sqrt{\sum_{i=1}^{H}\sum_{j=1}^{W}y(i,j,k)^2}}\right \}
\end{equation}
where $C$ represents the number of channels (colour components) in each pixel; $H$ represents the height of the image; $W$ represents the width of the image; $x(i,j,k)$ represents the pixel value of the $i$th row, $j$th column, and $k$th channel in the query image $X$; and $y(i,j,k)$ represents the pixel value of the $i$th row, $j$th column, and $k$th channel in the retrieving image $Y$.

\textbf{Universal Image Quality Index (UIQ)} takes into account the similarity between two images based on their correlation, luminance, and contrast. A higher \gls{UQI} value signifies a greater similarity between the two images. The maximum possible value of \gls{UQI} is 1, indicating that the two images are exactly the same.

\begin{equation}
   \mathrm{UIQ} = \frac{1}{C}\sum_{k=1}^{C}\frac{\sigma_{x_ky_k}}{\sigma_{x_k}\sigma_{y_k}}\cdot \frac{2 \overline{x_k}\overline{y_k}}{(\overline{x_k})^2 + (\overline{y_k})^2}\cdot \frac{2 \sigma_{x_k}\sigma_{y_k}}{\sigma_{x_k}^2 + \sigma_{y_k}^2}
\end{equation}
where $\overline{x_k}$ is the mean of the $k$th channel's pixel values in the query image $X$;
$\overline{y_k}$ is the mean of the $k$th channel's pixel values in the retrieving image $Y$;
$\sigma_{x_k}^2$ is the variance of the $k$th channel's pixel values in the query image $X$;
$\sigma_{y_k}^2$ is the variance of the $k$th channel's pixel values in the retrieving image $Y$;
$\sigma_{x_k}$ is the standard deviation of the $k$th channel's pixel values in the query image $X$;
$\sigma_{y_k}$ is the standard deviation of the $k$th channel's pixel values in the retrieving image $Y$;
$\sigma_{x_ky_k}$ is the correlation of the $k$th channel's pixel values between the query image $X$ and the retrieving image $Y$.


\subsubsection{Feature-based Similarity Metrics}
\label{sec:feasm}

\textbf{Euclidean distance} calculates the direct straight-line distance between each point of two vectors. A higher value of the Euclidean distance signifies a greater dissimilarity between the two images.

\begin{equation}
   \text{Euclidean Distance} = \sqrt{\sum_{i=1}^{N}(x_i-y_i)^2}
\end{equation}
where: $N$ represents the number of elements in the feature vector; $x_i$ represents the $i$th value of the extracted feature vector from the query image $X$; $y_i$ represents the $i$th value of the extracted feature vector from the retrieving image $Y$.\\

\textbf{Cosine distance} calculates the cosine of the angles between two vectors. A higher value of the cosine distance indicates that the two images are more similar.

\begin{equation}
   \text{Cosine Distance} = \frac{\sum_{i=1}^{N}x_i\cdot y_i}{ \sqrt{\sum_{i=1}^{N}x_i^2}\cdot \sqrt{\sum_{i=1}^{N}y_i^2}}
\end{equation}
where:\\
$N$ represents the number of elements in the feature vector; $x_i$ represents the $i$th value of the extracted feature vector from image $X$; and $y_i$ represents the $i$th value of the extracted feature vector from image $Y$.

\textbf{Manhattan distance} calculates the sum of absolute differences between corresponding elements of two vectors. A larger value of the Manhattan distance signifies that the two images are more dissimilar.

\begin{equation}
   \text{Manhattan Distance} = \sum_{i=1}^{N}|x_i - y_i|
\end{equation}
where:\\
$N$ represents the number of elements in the feature vector; $x_i$ represents the $i$th value of the extracted feature vector from the qeury image $X$; and $y_i$ represents the $i$th value of the extracted feature vector from the retrieving image $Y$.

\textbf{Jaccard distance} calculates the ratio of the common elements between two feature vectors to the total number of elements present in the vectors. A higher value of the Jaccard distance suggests that the two images are more similar in terms of the shared features or values.

\begin{equation}
   \text{Jaccard Distance} = \frac{x \cap y}{x \cup y}
\end{equation}
where:\\
$x \cap y$ represents the intersection of the sets of elements present in the feature vectors of the query image $X$ and retrieving image $Y$; and $x \cup y$ represents the union of the sets of elements present in the feature vectors of the query image $X$ and retrieving image $Y$.


\subsection{Evaluation Measure}
\label{sec:eval}
This section introduces the evaluation measures employed within the scope of an \gls{IR} task carried out in this experiment. In the context of an \gls{IR} task, the primary objective revolves around searching for images with similar irregular patterns within a datastore. $\mathrm{Precision}@K$ is defined as the ratio of relevant images with irregular patterns correctly retrieved among the top $K$ retrieved images with irregular patterns. Moreover, the $\mathrm{AP}@K$, \gls{AP}, calculates the average value of $\mathrm{Precision}@K$ across the queries in an irregular pattern class. Subsequently, the term $\mathrm{mAP}@K$, signifying mean average precision, computes the average of $\mathrm{AP}@K$ values across all the irregular pattern classes that exist within the dataset. Furthermore, the standard deviations for \gls{AP} and \gls{mAP} were computed to assess the consistency of the retrieval performance for query and class, respectively.

\subsection{Methdology}
\label{sec:metho}
The experiment ran separate \gls{IR} tasks to evaluate the performance between different features and similarity metrics for each dataset. Three conventional image-based similarity metrics (i.e.\ \gls{MSE}, \gls{SAM}, and \gls{UQI}) were employed in instances where the raw image data was utilised. For scenarios involving either \gls{DC} or \gls{LBP} feature data, the experiment utilised four feature-based similarity metrics (i.e.\ Euclidean, Cosine, Manhattan, and Jaccard). Notably, in alignment with Lowe's recommendations \cite{09_sift}, the Euclidean similarity metric was exclusively utilised for the \gls{SIFT} feature data.

In the initial step, this experiment compresses the raw images into four distinct sizes (i.e.\ $100\times 100$, $50\times 50$, $20\times 20$, $8\times 8$) with the dual goals of normalisation and acceleration of the retrieval process. Subsequently, the \gls{SIFT} and \gls{LBP} features are extracted from these compressed images. Notably, the extraction of \glspl{DC} necessitates the utilisation of raw images due to the potential distortion of mask annotations caused by image resizing. The next step iteratively picks the compressed image or the extracted feature vector of each irregular pattern in the dataset as a query; and then retrieves the remaining irregular patterns by applying the corresponding similarity metrics; for instance, the feature-based similarity metrics are utilised for \glspl{DC}, \gls{SIFT} and \gls{LBP} features and image-based similarity metrics are utilised for compressed raw images. The retrieval criterion for relevant irregular patterns is based on the class of the irregular pattern. Then, the retrieved irregular patterns are ranked according to the computed similarities, and the precision is computed for each query. Ultimately, the experiment calculates the average precision to evaluate the retrieval performance for each class within the dataset; the mean average precision is calculated to assess the overall retrieval performance.

The experiment was conducted on a high-performance computer featuring an AMD Ryzen 9 CPU and 32GB RAM. Notably, the utilisation of a GPU is unnecessary for executing the \gls{IR} task.

\section{Results \& Discussion}
\label{sec:res}

This section illustrates the retrieval performance results when using different extracted features and similarity metrics for each dataset. The full retrieval results are presented in Appendices \ref{app:chestct}, \ref{app:heatsink}, \ref{app:lakeice}, \ref{app:wtb}. In each appendix, a series of tables provide insight into the average precision with the associated standard deviation for each class within the dataset; additionally, there is a table illustrating the mean average precision with its standard deviation to reflect the overall performance in the dataset. This section contains two subsections to illustrate the evaluations of the retrieval performance. Section \ref{sec:res1} assesses the retrieval performance using distinct features and similarity metrics for each dataset. This assessment aims to identify noteworthy features and similarity metrics that exhibit impressive performance. Section \ref{sec:res2} compares the retrieval performance of the remarkable methods mentioned in Section \ref{sec:res1} for each dataset; moreover, this section discusses the time used to extract the features and retrieve the query for each dataset.

\subsection{Information Retrieval Performances between Different Features, Similarity Metrics and Image Sizes for Each Dataset}
\label{sec:res1}

This section empirically evaluates the retrieval performance using four different features (i.e., \glspl{DC}, resized raw images, \gls{LBP}, and \gls{SIFT}) for each dataset. The \glspl{DC}-based methods applied the \glspl{DC}, extracted from the raw images, to four feature-based similarity metrics (i.e.\ Cosine, Euclidean, Jaccard and Manhattan). The image-based methods applied the images with four different sizes (i.e.\ $8 \times 8$, $20 \times 20$, $50 \times 50$, $100 \times 100$) to three image-based similarity metrics (i.e.\ \gls{MSE}, \gls{UQI} and \gls{SAM}). The \gls{LBP}-based methods applied the \gls{LBP} features, extracted from resized images, to four feature-based similarity metrics. The \gls{SIFT}-based methods applied \gls{SIFT} features, extracted from resized images, to the Euclidean similarity metric. Then, this section discusses the noteworthy similarity metrics and image sizes that achieved an outstanding retrieval performance (e.g.\ highest \gls{mAP} and lowest standard deviation) when applying different features for each class within the dataset.

\subsubsection{Chest CT Dataset}
Table \ref{tab:chestctm} shows the \glspl{mAP}, along with standard deviation, when using different features, similarity metrics and image sizes in chest \gls{CT} dataset; Table \ref{tab:chestct1}, \ref{tab:chestct2}, \ref{tab:chestct3} illustrate the \glspl{AP} with standard deviation for each class within the dataset. 

\textbf{Performance using \glspl{DC}-based methods [Proposed]:} There were two similarity metrics (i.e.\ Cosine and Euclidean) that yielded the highest \gls{mAP} and the lowest standard deviation, averaging at 0.85 $\pm$ 0.06. In terms of retrieval performance for each class using the Cosine or Euclidean metric, both metrics had same \gls{mAP} values and standard deviations for class 1. For class 2, the Euclidean metric achieved a slightly higher \gls{AP} of 0.01 compared to the Cosine metric at $@1$ and $@20$. Conversely, the Cosine metric outperformed the Euclidean metric by 0.01 in \gls{AP} at $@1$ and $@10$ for class 3. Additionally, the Manhattan metric exhibited noteworthy retrieval performance for the Chest \gls{CT} dataset and relatively achieved the highest \gls{AP} for class 1 and 2. However, for class 3, the retrieval performance of the Manhattan metric was slightly lower than that of the Cosine and Euclidean metrics, with an average difference of 0.01-0.02 at $@5$, $@10$, and $@15$. This resulted in a higher standard deviation in \gls{mAP}, despite the metrics having the same mean \gls{mAP} values.

\textbf{Performance using image-based methods:} 

The best-performing image-based method was the \gls{UQI} metric with a $20 \times 20$ image size, averaging at 0.76 $\pm$ 0.17 in terms of \gls{mAP}. The \gls{UQI} metrics with $50 \times 50$ and $100 \times 100$ image sizes exhibited similar \gls{mAP} values across the range from $@1$ to $@20$, but they had higher standard deviations compared to the \gls{UQI} metric with a $20 \times 20$ image size. The \gls{UQI} metric with a $20 \times 20$ image size showed a relatively small difference between the maximum and minimum \gls{AP} in different classes (i.e.\ $@1$: 0.94 - $@20$: 0.89 for class 1 and $@1$: 0.59 - $@20$: 0.58 for class 3), although it may not achieve the highest \gls{AP}.

\textbf{Performance using \gls{LBP}-based methods:} The performance of the \gls{LBP}-based methods exhibited a correlation with the image size. The highest \gls{mAP} was achieved, averaging at 0.30 $\pm$ 0.22 between $@1$ and $@20$, when using images of size $100 \times 100$. However, it is worth noting that the retrieval performance of the best-performing \gls{LBP} method for each class was not consistent. 

\textbf{Performance using \gls{SIFT}-based methods:} The $20 \times 20$ image size proved to be the optimal setting for the \gls{SIFT}-based method, with an average \gls{mAP} of 0.52 $\pm$ 0.31. The standard deviation of the \gls{SIFT} method was the highest among all methods; consequently, the \gls{SIFT} method struggled to maintain consistent retrieval performance across all classes within the chest \gls{CT} dataset.

\subsubsection{Heatsink Dataset}

Table \ref{tab:heatsinkm} shows the \glspl{mAP}, along with standard deviation, when using different features, similarity metrics and image sizes in heatsink dataset; Table \ref{tab:heatsink1} and \ref{tab:heatsink2} illustrate the \glspl{AP} with standard deviation for each class within the dataset.

\textbf{Performance using \glspl{DC}-based methods [Proposed]:} All similarity metrics consistently achieved the highest \gls{mAP} with an average of 0.97 $\pm$ 0.02, except for the Jaccard metric. When analysing the performance for each class, the Manhattan metric sometimes exhibited a slightly higher standard deviation by 0.01, compared to the Cosine and Euclidean metrics. Additionally, the \gls{AP} for the Manhattan metric was sometimes lower by 0.01. As a result, the performance of the Manhattan metrics was slightly lower than that of the others, particularly at $@10$ and $@15$.

\textbf{Performance using image-based methods:} The average \gls{mAP} between $@1$ and $@20$ reached its peak at 0.88 $\pm$ 0.08 when using $100 \times 100$ images with the \gls{UQI} metric, although its $\mathrm{mAP}@1$ was slightly lower at 0.86 compared to others. Additionally, \gls{MSE} with $8 \times 8$ image size, demonstrated relatively high performance with an average \gls{mAP} of 0.87. When considering the performance for each class, the \gls{UQI} metric with $100 \times 100$ image size and the \gls{MSE} metric with $8 \times 8$ image size consistently maintained a high \gls{AP} across all classes. In contrast, other methods exhibited fluctuations in \gls{AP} when applied to different classes.

\textbf{Performance using \gls{LBP}-based methods:} The \gls{LBP} method exhibited consistent \gls{mAP} and standard deviation across all feature-based similarity metrics. The best-performing \gls{LBP} method achieved an \gls{mAP} of 0.53 $\pm$ 0.18 on average when utilising $8 \times 8$ images. However, it is worth noting that the performance of the \gls{LBP}-based method varied significantly for each class. The \gls{LBP} metric with $8 \times 8$ images significantly outperformed the others by more than 0.43 in \gls{AP} for class 2; but, the performance of this metric fell behind the others for class 1, particularly when compared to the \gls{LBP} metric with a $20 \times 20$ image size.

\textbf{Performance using \gls{SIFT}-based methods:} The best retrieval performance was 0.54 $\pm$ 0.23 in terms of \gls{mAP} when applying the $100 \times 100$ images to the \gls{SIFT}-based method. The performance of the best-performing \gls{SIFT} method was not balanced between each class; for instance, the \gls{AP} ranged from $@5$: 0.51 to $@20$: 0.45 for class 1 and $@5$: 0.59 to $@20$: 0.64 for class 2.

\subsubsection{Lake Ice Dataset}

Table \ref{tab:lakeicem} shows the \glspl{mAP}, along with standard deviation, when using different features, similarity metrics and image sizes in lake ice dataset; Table \ref{tab:lakeice1}, \ref{tab:lakeice2}, \ref{tab:lakeice3} and \ref{tab:lakeice4} illustrate the \glspl{AP} with standard deviation for each class within the dataset. 

\textbf{Performance using \glspl{DC}-based methods [Proposed]:} The highest \gls{mAP} among the \gls{DC}-based methods had an average of 0.90 $\pm$ 0.07 when using the Manhattan metric. Also, the retrieval performance was relatively balanced across all classes within the dataset; the \gls{AP} for all classes exceeded 0.94 at $@1$ and 0.74 at $@20$, which was higher than for other \gls{DC}-based methods.

\textbf{Performance using image-based methods:} All image-based methods showed similar performance; however, the \gls{SAM} using $8 \times 8$ images outperformed the others with the highest \gls{mAP} (0.86) and the lowest standard deviation (0.13). In terms of performance across different classes, there was no significant difference observed between all similarity metrics and image sizes. Nevertheless, the \gls{SAM} with $8 \times 8$ image size achieved a higher \gls{AP} than other image-based methods by 0.04-0.06 for class 4, resulting in a higher \gls{mAP} with a lower standard deviation.

\textbf{Performance using \gls{LBP}-based methods:} All \gls{LBP}-based methods exhibited relatively low \gls{mAP} values across all classes. The best-performing \gls{LBP}-based method, when using any feature-based similarity metric with $100 \times 100$ images, only reached an average \gls{mAP} of 0.26 $\pm$ 0.37. Furthermore, this best-performing \gls{LBP}-based method achieved high performance for class 3 (from $\mathrm{mAP}@1$: 0.99 to $\mathrm{mAP}@20$: 0.68); for other classes, however, the \gls{mAP} dropped significantly, falling below 0.23 and even reaching 0.00.

\textbf{Performance using \gls{SIFT}-based methods:} In all \gls{SIFT}-based methods, using large-sized images (i.e., $100 \times 100$) achieved the highest \gls{mAP}, averaging at 0.64 $\pm$ 0.15. The performance of the \gls{SIFT} method with a $100 \times 100$ image size was relatively balanced between all classes, except for class 4, where the \gls{AP} was \SI{40}{\percent} lower than that for other classes.

\subsubsection{Wind Turbine Blade Dataset}

Table \ref{tab:wtbm} shows the \glspl{mAP}, along with standard deviation, when using different features, similarity metrics and image sizes in wind turbine blade dataset; Table \ref{tab:wtb1}, \ref{tab:wtb2}, \ref{tab:wtb3} and \ref{tab:wtb4} illustrate the \glspl{AP} with standard deviation for each class within the dataset. 

\textbf{Performance using \glspl{DC}-based methods [Proposed]:} In the \glspl{DC}-based methods, the Manhattan metric outperformed others with the highest \gls{mAP} (0.62) and the lowest standard deviation (0.17). When considering the performance for each class, the retrieval performance using the Manhattan metric generally exceeded other metrics, except for $\mathrm{AP}@15$ and $\mathrm{AP}@20$ in class 2, and $\mathrm{AP}@5$, $\mathrm{AP}@10$, and $\mathrm{AP}@15$ in class 4.

\textbf{Performance using image-based methods:} In the image-based methods, two settings (i.e., \gls{MSE} with a $8 \times 8$ image size and \gls{UQI} with a $20 \times 20$ image size) both averagely achieved the highest \gls{mAP} (0.44) and the lowest standard deviation (0.31). When evaluating the performance for each class using these best-performing image-based methods, they exhibited similar \gls{AP} values across all classes. However, the \gls{AP} of the \gls{MSE} metric occasionally exceeded that of the \gls{UQI} by 0.03-0.08 at $@1$ for all classes except class 3.

\textbf{Performance using \gls{LBP}-based methods:} Any feature-based similarity metric with a small-sized image (i.e.\ $8 \times 8$) outperformed all \gls{LBP}-based methods and achieved an average \gls{mAP} of 0.26 $\pm$ 0.17. However, all \gls{LBP}-based methods, including the best-performing one, struggled to maintain a consistent performance across all classes. For instance, the \gls{AP} of the best-performing method was lower than other \gls{LBP}-based methods between $@1$ and $@20$ for classes 1 and 3. 

\textbf{Performance using \gls{SIFT}-based methods:} The highest \gls{mAP} was achieved at 0.35 $\pm$ 0.30 when utilising the Euclidean metric with a $100 \times 100$ image size in the \gls{SIFT}-based method. When evaluating the performance differences of the \gls{SIFT} methods across each class, the best-performing method was relatively more accurate than others for class 3 by over 0.10, although it did not outperform the others for the rest of the classes.

\subsection{Overall Performance Comparisons for the Information Retrieval Tasks}
\label{sec:res2}
This section compares retrieval performance of different feature extraction methods, utilising the settings that demonstrated remarkable results for each dataset. These settings were selected based on the analysis presented in Section \ref{sec:res1}. Additionally, this section delves into investigating the time completed for each method and dataset in the \gls{IR} task.

\begin{figure}[!ht]
\centering
\includegraphics[width=\linewidth]{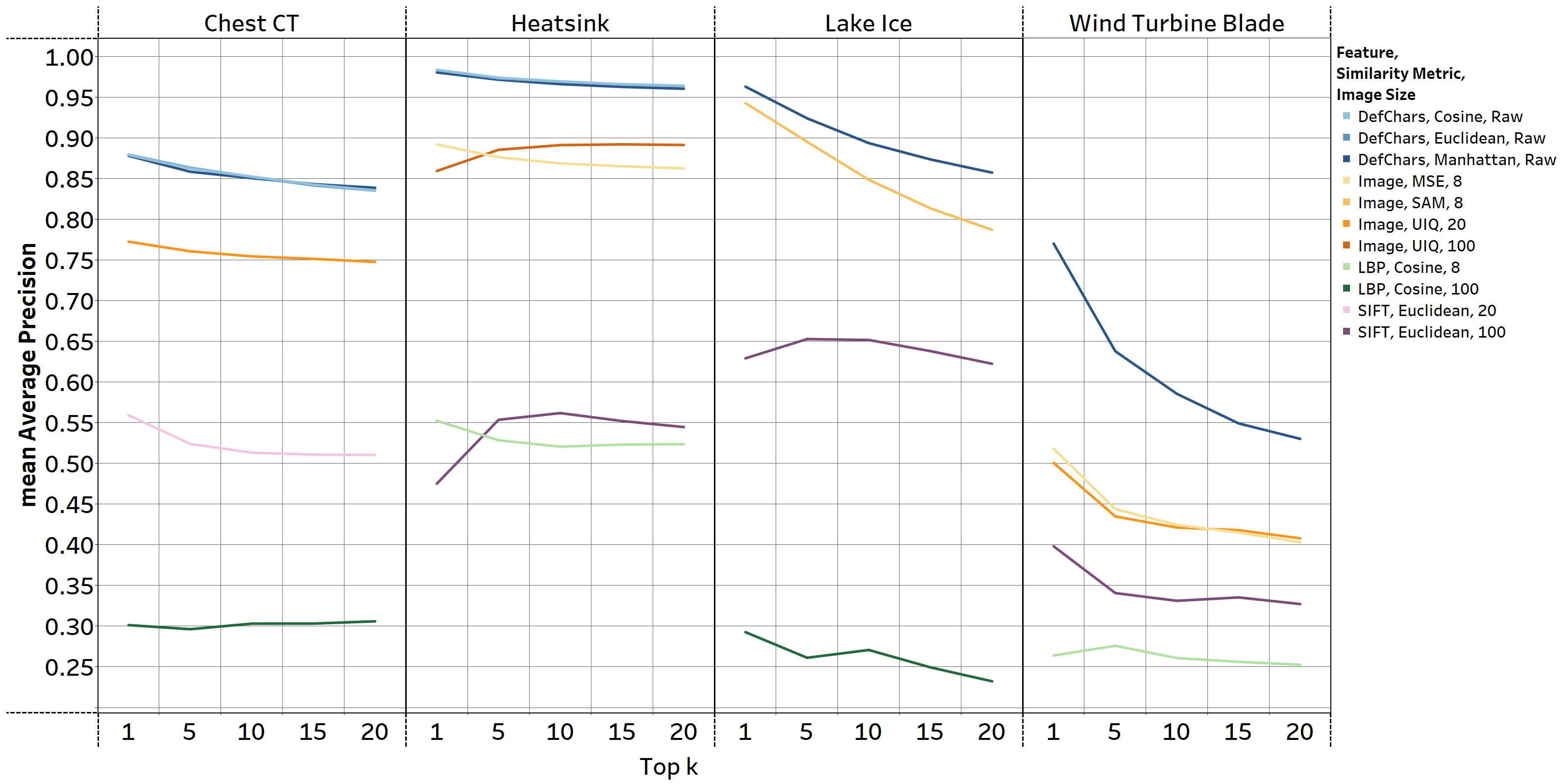}

\caption{Mean Average Precision of the highlighted features, similarity metrics and image sizes. The blue-based, orange-based, green-based, and purple-based lines respectively represent the \gls{IR} methods which used the \glspl{DC}, image-based feature, \gls{LBP} feature, and \gls{SIFT} feature. The depths of the line colour represent different similarity metrics and image sizes.}
\label{fig:map_sel}
\end{figure} 

Figure \ref{fig:map_sel} presents a line chart illustrating the $\mathrm{mAP}$ of the outstanding \gls{IR} methods, that were explained earlier, across all datasets. The \glspl{DC}-based methods consistently outperformed other methods in all datasets. When using \glspl{DC}, the performances of different similarity metrics were not notably distinct; however, the Manhattan metric stood out as an effective choice because it consistently reached the highest \gls{mAP} across all datasets. The image-based \gls{IR} methods were the second best method in all datasets. However, the choice of similarity metrics and image sizes emerged as significant factors influencing performance for different datasets, and there is no image-based method with a consistent setting that relatively maintains high retrieval performance. For instance, the \gls{UQI} metric performed relatively better with large-sized images in the chest CT and heatsink datasets, while the \gls{SAM} and \gls{MSE} metrics achieved higher \gls{mAP} in lake ice and wind turbine blade dataset when utilising small-sized images. The performance of the \gls{LBP}-based method was comparably the worst in all remarkable methods and did not exhibit significant differences based on the choice of similarity metric. Furthermore, no discernible pattern emerged regarding the impact of image sizes on \gls{LBP} methods. For instance, the chest \gls{CT} and lake ice datasets exhibited higher $\mathrm{mAP}$ values with larger-sized images, whereas the heatsink and wind turbine blade datasets yielded better results with smaller-sized images. The \gls{SIFT}-based methods exhibited a better overall performance compared to the \gls{LBP}-based methods, although they fell short of other methods. Notably, for the \gls{SIFT} methods, larger image sizes (e.g.\ $100 \times 100$) translated to enhanced performance compared to smaller sizes except for the chest \gls{CT} dataset.

\begin{figure}[!ht]
\centering
\includegraphics[width=\linewidth]{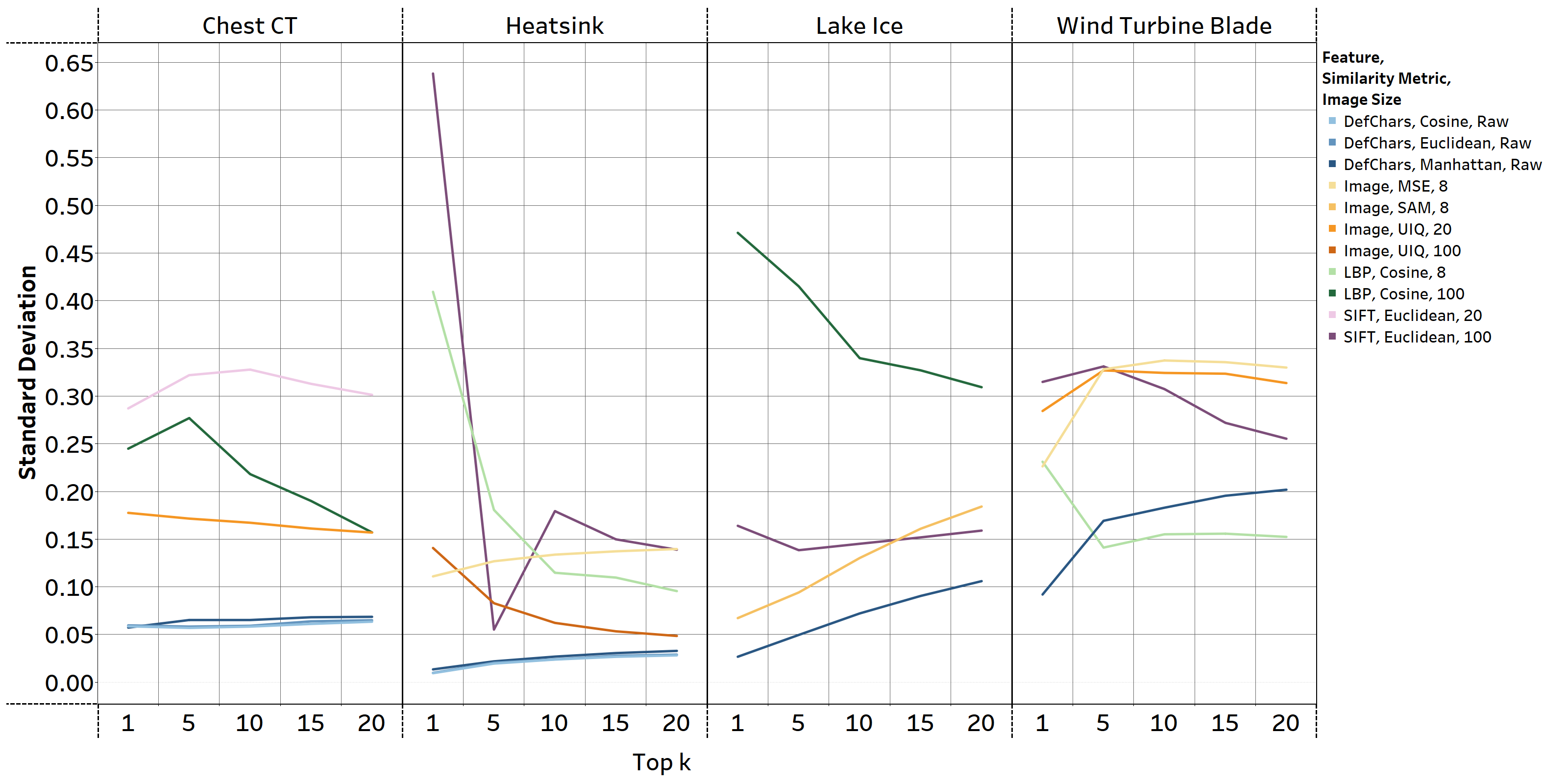}

\caption{Standard deviation of the highlighted features, similarity metrics and image sizes. The blue-based, orange-based, green-based, and purple-based lines respectively represent the \gls{IR} methods which used the \glspl{DC}, image-based feature, \gls{LBP} feature, and \gls{SIFT} feature. The depths of the line colour represent different similarity metrics and image sizes.}

\label{fig:sd_sel}
\end{figure} 

Figure \ref{fig:sd_sel} provides insight into the standard deviations encountered when calculating the $\mathrm{mAP}$ using the outstanding methods. Notably, the \glspl{DC} methods generally exhibited the lowest standard deviation across all datasets. This indicates their relatively stable and reliable performance. In the wind turbine blade dataset, the \gls{LBP} method displayed standard deviations that were 0.03-0.05 lower than those of the \gls{DC} method when retrieving more than 5 irregular patterns. However, the $\mathrm{mAP}$ achieved by the \gls{LBP} method was the lowest among all methods. In contrast, the rest of the methods, such as image-based, \gls{SIFT}-based, and \gls{LBP}-based methods, exhibited higher standard deviations, often exceeding those of \glspl{DC}-based methods by more than 0.1 in all datasets. Moreover, these methods demonstrated varying standard deviations across different datasets. This implies that these methods might exhibit less consistency when retrieving irregular patterns from different classes within a dataset. As noted in Table \ref{tab:typedist}, the datasets themselves are relatively imbalanced between each class. However, the \gls{IR} method utilising \glspl{DC} and the Manhattan metric showed the capability to maintain relatively high and balanced accuracy in retrieving similar irregular patterns. Additionally, the performance of \glspl{DC} methods did not exhibit significant deterioration when applied to a small dataset (i.e.\ wind turbine blade).

\begin{figure}[!ht]
\includegraphics[width=\linewidth]{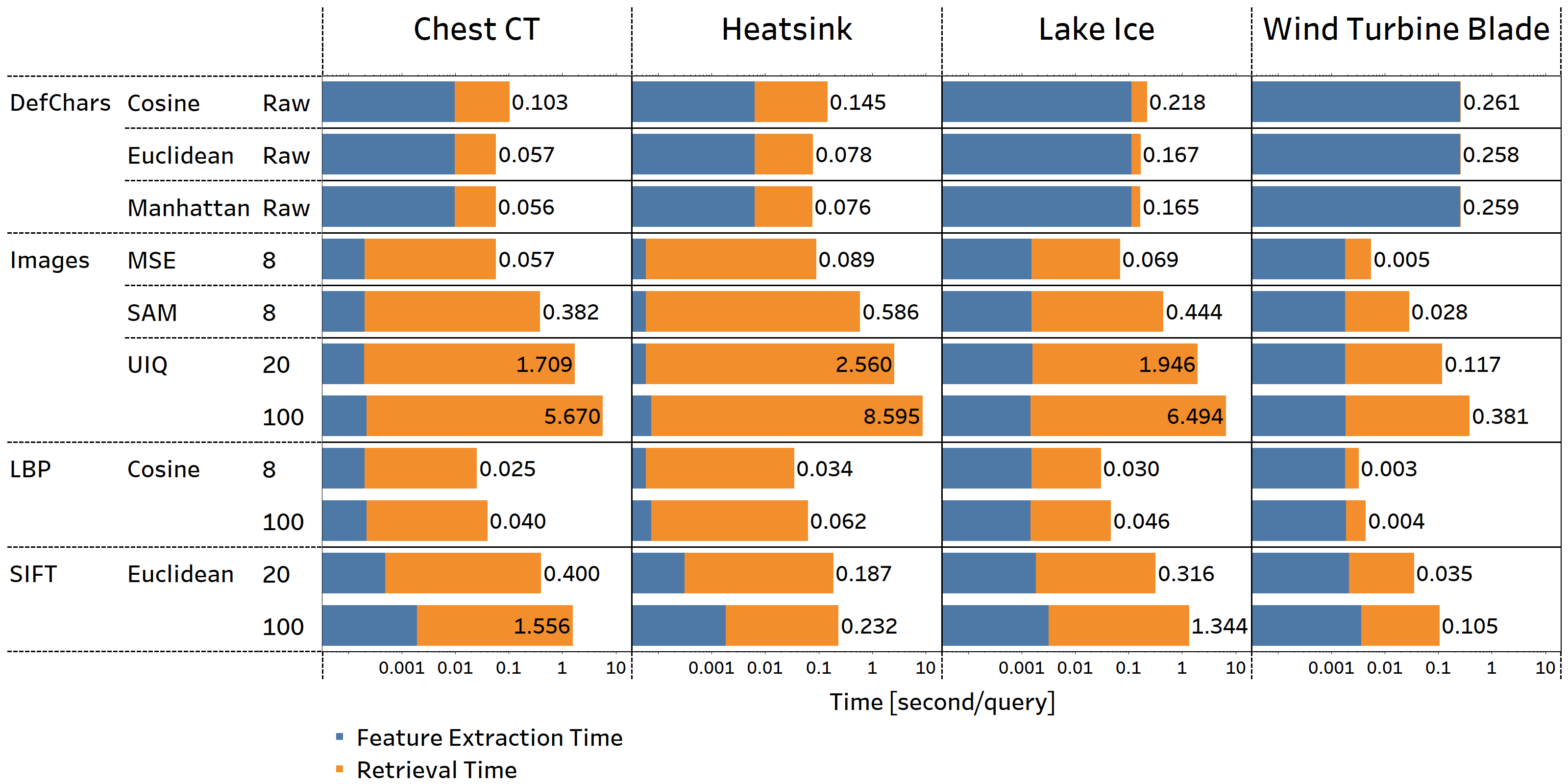}
\caption{Image Retrieval Time of the highlighted features, similarity metrics and image size for each dataset. The blue bar represents the average feature extraction time and the orange bar represents the average retrieval time. The text at end of each bar shows the total time used in retrieving each query.}
\label{fig:time_sel}
\end{figure} 

Figure \ref{fig:time_sel} presents the time required to extract features and retrieve images with similar irregular patterns for each query across four datasets. Among these remarkable methods, \gls{LBP} exhibited the shortest total time, taking less than 0.062 seconds per query. On the other hand, \gls{UQI} consumed the most time, especially when using $100 \times 100$ images, resulting in query times exceeding 5 seconds for large datasets such as chest \gls{CT}, heatsink, and lake ice. In the \glspl{DC}-based method, the retrieval time ranged from 0.06 to 0.26 seconds, making it one of the faster approaches except for the wind turbine blade dataset. However, while the feature extraction time for \glspl{DC}-based method was comparatively longer than other methods, their retrieval times were significantly shorter than those of other methods. The image-based approaches generally required more time for retrieval, although image resizing reduced the feature extraction time. The \gls{SIFT}-based method needed an additional 0.0003 to 0.0018 seconds for extracting the \gls{SIFT} features after image resizing, but the retrieval time was relatively shorter than most image-based methods. When considering the time impact from the dataset or image size, a large dataset or image size typically led to a longer execution time. For instance, the \gls{IR} time for \gls{UQI} metrics when using $100 \times 100$ images was over three times longer than when using $20 \times 20$ images. On the other hand, the heatsink dataset, the largest dataset according to Table \ref{tab:typedist}, contained 7007 irregular patterns, resulting in longer retrieval times across all datasets. Therefore, the \glspl{DC}-based method was able to complete a relatively fast and accurate \gls{IR} task.


\section{Conclusions \& Future Work}
\label{sec:conc}

This paper proposed an \gls{IR} framework to retrieve images with irregular patterns and completed a comprehensive evaluation of retrieval performance using different feature extraction methods (i.e.\ \glspl{DC}, resized raw images, \gls{LBP}, and \gls{SIFT}) along with different similarity metrics across four datasets (chest \gls{CT}, heatsink, lake ice, and wind turbine blade). The findings highlighted that the \gls{IR} framework, utilising \glspl{DC} and the Manhattan similarity metric, consistently demonstrated robust and high performance with a relatively fast \gls{IR} speed across all datasets. Moreover, this method did not exhibit a significant bias toward each class within the dataset, despite minor fluctuations in $\mathrm{AP}$ values possibly attributed to dataset complexity. Future work could involve comparing the \glspl{DC}-based method with retrieval performance using deep learning-based features or alternative similarity metrics, investigating explanations for the \gls{IR} task, such as plotting the visualised charts of the \glspl{DC}, and utilising segmentation or object detection techniques to automatically complete annotations of the irregular patterns for the \gls{IR} task.


\bibliographystyle{IEEEtran}
\bibliography{references}  

\begin{thebibliography}{10}
\providecommand{\url}[1]{#1}
\csname url@samestyle\endcsname
\providecommand{\newblock}{\relax}
\providecommand{\bibinfo}[2]{#2}
\providecommand{\BIBentrySTDinterwordspacing}{\spaceskip=0pt\relax}
\providecommand{\BIBentryALTinterwordstretchfactor}{4}
\providecommand{\BIBentryALTinterwordspacing}{\spaceskip=\fontdimen2\font plus
\BIBentryALTinterwordstretchfactor\fontdimen3\font minus \fontdimen4\font\relax}
\providecommand{\BIBforeignlanguage}[2]{{%
\expandafter\ifx\csname l@#1\endcsname\relax
\typeout{** WARNING: IEEEtran.bst: No hyphenation pattern has been}%
\typeout{** loaded for the language `#1'. Using the pattern for}%
\typeout{** the default language instead.}%
\else
\language=\csname l@#1\endcsname
\fi
#2}}
\providecommand{\BIBdecl}{\relax}
\BIBdecl

\bibitem{01_wafer}
T.~Nakazawa and D.~V. Kulkarni, ``{Wafer Map Defect Pattern Classification and Image Retrieval Using Convolutional Neural Network},'' \emph{IEEE Transactions on Semiconductor Manufacturing}, vol.~31, no.~2, pp. 309--314, 2018.

\bibitem{02_fabric}
\BIBentryALTinterwordspacing
X.~Hu, M.~Fu, Z.~Zhu, Z.~Xiang, M.~Qian, and J.~Wang, ``{Unsupervised defect detection algorithm for printed fabrics using content-based image retrieval techniques},'' \emph{Textile Research Journal}, vol.~91, no. 21-22, pp. 2551--2566, 2021. [Online]. Available: \url{https://doi.org/10.1177/00405175211008614}
\BIBentrySTDinterwordspacing

\bibitem{04_bridge}
P.~Liu and N.~El-Gohary, ``{Semantic Image Retrieval and Clustering for Supporting Domain-Specific Bridge Component and Defect Classification},'' in \emph{Construction Research Congress 2020}, 11 2020, pp. 809--818.

\bibitem{05_chestxray}
S.~Agrawal, A.~Chowdhary, S.~Agarwala, V.~Mayya, and S.~Kamath~S, ``Content-based medical image retrieval system for lung diseases using deep cnns,'' \emph{International Journal of Information Technology}, vol.~14, no.~7, pp. 3619--3627, 2022.

\bibitem{20_ct}
B.~Xie, Y.~Zhuang, N.~Jiang, and J.~Liu, ``{An effective and efficient framework of content-based similarity retrieval of large {CT} image sequences based on {WSLEN} model},'' \emph{Multimedia Tools and Applications}, 09 2023.

\bibitem{21_lung}
\BIBentryALTinterwordspacing
J.~Choe, H.~J. Hwang, J.~B. Seo, S.~M. Lee, J.~Yun, M.-J. Kim, J.~Jeong, Y.~Lee, K.~Jin, R.~Park, J.~Kim, H.~Jeon, N.~Kim, J.~Yi, D.~Yu, and B.~Kim, ``{Content-based Image Retrieval by Using Deep Learning for Interstitial Lung Disease Diagnosis with Chest CT},'' \emph{Radiology}, vol. 302, no.~1, pp. 187--197, 2022, pMID: 34636634. [Online]. Available: \url{https://doi.org/10.1148/radiol.2021204164}
\BIBentrySTDinterwordspacing

\bibitem{ex1_ice}
\BIBentryALTinterwordspacing
K.~A. Scott, L.~Xu, and H.~K. Pour, ``{Retrieval of ice/water observations from synthetic aperture radar imagery for use in lake ice data assimilation},'' \emph{Journal of Great Lakes Research}, vol.~46, no.~6, pp. 1521--1532, 2020. [Online]. Available: \url{https://www.sciencedirect.com/science/article/pii/S0380133020301994}
\BIBentrySTDinterwordspacing

\bibitem{ex2_ice}
\BIBentryALTinterwordspacing
E.~Stonevicius, G.~Uselis, and D.~Grendaite, ``{Ice Detection with Sentinel-1 SAR Backscatter Threshold in Long Sections of Temperate Climate Rivers},'' \emph{Remote Sensing}, vol.~14, no.~7, p. 1627, Mar 2022. [Online]. Available: \url{http://dx.doi.org/10.3390/rs14071627}
\BIBentrySTDinterwordspacing

\bibitem{11_uqi}
Z.~Wang and A.~Bovik, ``{A universal image quality index},'' \emph{IEEE Signal Processing Letters}, vol.~9, no.~3, pp. 81--84, 2002.

\bibitem{12_sam}
R.~H. Yuhas, A.~F. Goetz, and J.~W. Boardman, ``{Discrimination among semi-arid landscape endmembers using the spectral angle mapper (SAM) algorithm},'' in \emph{JPL, Summaries of the Third Annual JPL Airborne Geoscience Workshop. Volume 1: AVIRIS Workshop}, 1992.

\bibitem{29_mse_sim}
T.~{VenkatNarayanaRao} and A.~{Govardhan}, ``{Assessment of Diverse Quality Metrics for Medical Images Including Mammography},'' \emph{International Journal of Computer Applications}, vol.~83, no.~4, pp. 42--47, 12 2013.

\bibitem{13_ssim}
Z.~Wang, A.~Bovik, H.~Sheikh, and E.~Simoncelli, ``{Image quality assessment: from error visibility to structural similarity},'' \emph{IEEE Transactions on Image Processing}, vol.~13, no.~4, pp. 600--612, 2004.

\bibitem{30_ct_sim}
B.~Rajith, M.~Srivastava, and S.~Agarwal, ``Edge preserved de-noising method for medical x-ray images using wavelet packet transformation,'' in \emph{Emerging Research in Computing, Information, Communication and Applications}, N.~R. Shetty, N.~Prasad, and N.~Nalini, Eds.\hskip 1em plus 0.5em minus 0.4em\relax New Delhi: Springer India, 2016, pp. 449--467.

\bibitem{31_sim}
Y.~Zhang, ``{Methods for image fusion quality assessment–A review, comparison and analysis},'' \emph{The International Archives of the Photogrammetry, Remote Sensing and Spatial Information Sciences}, vol.~37, 01 2008.

\bibitem{22_wlbp_steeldefect}
F.~Z. Boudani, N.~Nacereddine, and N.~Laiche, ``"content-based image retrieval for surface defects of hot rolled steel strip using wavelet-based lbp",'' in \emph{Progress in Artificial Intelligence and Pattern Recognition}, Y.~Hern{\'a}ndez~Heredia, V.~Mili{\'a}n~N{\'u}{\~{n}}ez, and J.~Ruiz~Shulcloper, Eds.\hskip 1em plus 0.5em minus 0.4em\relax Cham: Springer International Publishing, 2021, pp. 404--413.

\bibitem{32_lbp_fea}
\BIBentryALTinterwordspacing
L.~Zhang, X.~Liu, Z.~Lu, F.~Liu, and R.~Hong, ``{Lace Fabric Image Retrieval Based on Multi-Scale and Rotation Invariant LBP},'' in \emph{Proceedings of the 7th International Conference on Internet Multimedia Computing and Service}, ser. ICIMCS '15.\hskip 1em plus 0.5em minus 0.4em\relax New York, NY, USA: Association for Computing Machinery, 2015. [Online]. Available: \url{https://doi.org/10.1145/2808492.2808567}
\BIBentrySTDinterwordspacing

\bibitem{33_lbp_fea}
A.~Khan, M.~H. Rajvee, B.~L. Deekshatulu, and L.~Pratap~Reddy, ``"a fused lbp texture descriptor-based image retrieval system",'' in \emph{Advances in Signal Processing, Embedded Systems and IoT}, V.~Chakravarthy, V.~Bhateja, W.~Flores~Fuentes, J.~Anguera, and K.~P. Vasavi, Eds.\hskip 1em plus 0.5em minus 0.4em\relax Singapore: Springer Nature Singapore, 2023, pp. 145--154.

\bibitem{34_lbp_fea}
\BIBentryALTinterwordspacing
A.~K., R.~S., K.~C., W.-C. Lai, S.~R. Srividhya, and N.~K., ``{A Modified LBP Operator-Based Optimized Fuzzy Art Map Medical Image Retrieval System for Disease Diagnosis and Prediction},'' \emph{Biomedicines}, vol.~10, no.~10, 2022. [Online]. Available: \url{https://www.mdpi.com/2227-9059/10/10/2438}
\BIBentrySTDinterwordspacing

\bibitem{35_sift_fea}
L.-j. Zhi, S.-m. Zhang, D.-z. Zhao, H.~Zhao, and S.-k. Lin, ``{Medical Image Retrieval Using SIFT Feature},'' in \emph{2009 2nd International Congress on Image and Signal Processing}, 2009, pp. 1--4.

\bibitem{36_sift_fea}
B.~F. Cruz, J.~T. de~Assis, V.~V. Estrela, and A.~Khelassi, ``A compact sift-based strategy for visual information retrieval in large image databases: Array,'' \emph{Medical Technologies Journal}, vol.~3, no.~2, pp. 402--412, 2019.

\bibitem{37_sift_fea}
\BIBentryALTinterwordspacing
M.~Srinivas, R.~R. Naidu, C.~Sastry, and C.~K. Mohan, ``{Content based medical image retrieval using dictionary learning},'' \emph{Neurocomputing}, vol. 168, pp. 880--895, 2015. [Online]. Available: \url{https://www.sciencedirect.com/science/article/pii/S0925231215006967}
\BIBentrySTDinterwordspacing

\bibitem{16_simfea1}
B.~Patel, k.~Yadav, and D.~Ghosh, ``{State-of-Art: Similarity Assessment for Content Based Image Retrieval System},'' in \emph{2020 IEEE International Symposium on Sustainable Energy, Signal Processing and Cyber Security (iSSSC)}, 2020, pp. 1--6.

\bibitem{27_fesim1}
\BIBentryALTinterwordspacing
K.~Seetharaman and S.~Sathiamoorthy, ``{A unified learning framework for content based medical image retrieval using a statistical model},'' \emph{Journal of King Saud University - Computer and Information Sciences}, vol.~28, no.~1, pp. 110--124, 2016. [Online]. Available: \url{https://www.sciencedirect.com/science/article/pii/S1319157815000889}
\BIBentrySTDinterwordspacing

\bibitem{08_morfea}
J.~{Zhang}, G.~{Cosma}, S.~{Bugby}, A.~{Finke}, and J.~{Watkins}, ``{Morphological Image Analysis and Feature Extraction for Reasoning with AI-based Defect Detection and Classification Models},'' \emph{arXiv e-prints}, p. arXiv:2307.11643, 7 2023.

\bibitem{07_fereview}
A.~Latif, A.~Rasheed, U.~Sajid, J.~Ahmed, N.~Ali, N.~I. Ratyal, B.~Zafar, S.~H. Dar, M.~Sajid, and T.~Khalil, ``{Content-Based Image Retrieval and Feature Extraction: A Comprehensive Review},'' \emph{Mathematical Problems in Engineering}, vol. 2019, pp. 1--21, August 2019.

\bibitem{color1}
H.~Shao, Y.~Wu, W.~Cui, and J.~Zhang, ``{Image Retrieval Based on MPEG-7 Dominant Color Descriptor},'' in \emph{2008 The 9th International Conference for Young Computer Scientists}, 2008, pp. 753--757.

\bibitem{color2}
X.~Duanmu, ``{Image Retrieval Using Color Moment Invariant},'' in \emph{2010 Seventh International Conference on Information Technology: New Generations}, 2010, pp. 200--203.

\bibitem{color3}
X.-Y. Wang, B.-B. Zhang, and H.-Y. Yang, ``{Content-based image retrieval by integrating color and texture features},'' \emph{Multimedia tools and applications}, vol.~68, no.~3, pp. 545--569, 2014.

\bibitem{color4}
\BIBentryALTinterwordspacing
Y.~Liu, D.~Zhang, and G.~Lu, ``{Region-based image retrieval with high-level semantics using decision tree learning},'' \emph{Pattern Recognition}, vol.~41, no.~8, pp. 2554--2570, 2008. [Online]. Available: \url{https://www.sciencedirect.com/science/article/pii/S0031320307005316}
\BIBentrySTDinterwordspacing

\bibitem{color5}
H.~Zhang, Z.~Dong, and H.~Shu, ``{Object recognition by a complete set of pseudo-Zernike moment invariants},'' in \emph{2010 IEEE International Conference on Acoustics, Speech and Signal Processing}, 2010, pp. 930--933.

\bibitem{color6}
J.-M. Guo, H.~Prasetyo, and J.-H. Chen, ``{Content-Based Image Retrieval Using Error Diffusion Block Truncation Coding Features},'' \emph{IEEE Transactions on Circuits and Systems for Video Technology}, vol.~25, no.~3, pp. 466--481, 2015.

\bibitem{color7}
\BIBentryALTinterwordspacing
Z.~Jiexian, L.~Xiupeng, and F.~Yu, ``{Multiscale Distance Coherence Vector Algorithm for Content-Based Image Retrieval},'' \emph{The Scientific World Journal}, vol. 2014, pp. 1--13, 2014. [Online]. Available: \url{https://doi.org/10.1155/2014/615973}
\BIBentrySTDinterwordspacing

\bibitem{color8}
M.~M. Islam, D.~Zhang, and G.~Lu, ``{Automatic Categorization of Image Regions Using Dominant Color Based Vector Quantization},'' in \emph{2008 Digital Image Computing: Techniques and Applications}, 2008, pp. 191--198.

\bibitem{texture1}
\BIBentryALTinterwordspacing
G.~Papakostas, D.~Koulouriotis, and V.~Tourassis, ``{Feature Extraction Based on Wavelet Moments and Moment Invariants in Machine Vision Systems},'' in \emph{Human-Centric Machine Vision}, M.~Chessa, F.~Solari, and S.~P. Sabatini, Eds.\hskip 1em plus 0.5em minus 0.4em\relax Rijeka: IntechOpen, 2012, ch.~2. [Online]. Available: \url{https://doi.org/10.5772/33141}
\BIBentrySTDinterwordspacing

\bibitem{texture2}
\BIBentryALTinterwordspacing
G.-H. Liu, Z.-Y. Li, L.~Zhang, and Y.~Xu, ``{Image retrieval based on micro-structure descriptor},'' \emph{Pattern Recognition}, vol.~44, no.~9, pp. 2123--2133, 2011, computer Analysis of Images and Patterns. [Online]. Available: \url{https://www.sciencedirect.com/science/article/pii/S0031320311000501}
\BIBentrySTDinterwordspacing

\bibitem{texture3}
\BIBentryALTinterwordspacing
X.~yuan Wang, Z.~feng Chen, and J.~jiao Yun, ``{An effective method for color image retrieval based on texture},'' \emph{Computer Standards \& Interfaces}, vol.~34, no.~1, pp. 31--35, 2012. [Online]. Available: \url{https://www.sciencedirect.com/science/article/pii/S0920548911000547}
\BIBentrySTDinterwordspacing

\bibitem{texture4}
\BIBentryALTinterwordspacing
R.~Ashraf, K.~Bashir, A.~Irtaza, and M.~T. Mahmood, ``{Content Based Image Retrieval Using Embedded Neural Networks with Bandletized Regions},'' \emph{Entropy}, vol.~17, no.~6, pp. 3552--3580, 2015. [Online]. Available: \url{https://www.mdpi.com/1099-4300/17/6/3552}
\BIBentrySTDinterwordspacing

\bibitem{texture5}
A.~Irtaza and M.~A. Jaffar, ``{Categorical image retrieval through genetically optimized support vector machines ({GOSVM}) and hybrid texture features},'' \emph{Signal, Image and Video Processing}, vol.~9, no.~7, pp. 1503--1519, Oct 2015.

\bibitem{texture6}
\BIBentryALTinterwordspacing
S.~Fadaei, R.~Amirfattahi, and M.~R. Ahmadzadeh, ``{Local derivative radial patterns: A new texture descriptor for content-based image retrieval},'' \emph{Signal Processing}, vol. 137, pp. 274--286, 2017. [Online]. Available: \url{https://www.sciencedirect.com/science/article/pii/S0165168417300695}
\BIBentrySTDinterwordspacing

\bibitem{texture7}
\BIBentryALTinterwordspacing
X.~Wang and Z.~Wang, ``{A novel method for image retrieval based on structure elements’ descriptor},'' \emph{Journal of Visual Communication and Image Representation}, vol.~24, no.~1, pp. 63--74, 2013. [Online]. Available: \url{https://www.sciencedirect.com/science/article/pii/S1047320312001605}
\BIBentrySTDinterwordspacing

\bibitem{spitial1}
\BIBentryALTinterwordspacing
N.~Ali, K.~B. Bajwa, R.~Sablatnig, S.~A. Chatzichristofis, Z.~Iqbal, M.~Rashid, and H.~A. Habib, ``{A Novel Image Retrieval Based on Visual Words Integration of SIFT and SURF},'' \emph{PLOS ONE}, vol.~11, no.~6, pp. 1--20, 06 2016. [Online]. Available: \url{https://doi.org/10.1371/journal.pone.0157428}
\BIBentrySTDinterwordspacing

\bibitem{spitial2}
S.~Lazebnik, C.~Schmid, and J.~Ponce, ``{Beyond Bags of Features: Spatial Pyramid Matching for Recognizing Natural Scene Categories},'' in \emph{2006 IEEE Computer Society Conference on Computer Vision and Pattern Recognition (CVPR'06)}, vol.~2, 2006, pp. 2169--2178.

\bibitem{spitial3}
Z.~Mehmood, S.~M. Anwar, N.~Ali, H.~A. Habib, and M.~Rashid, ``{A Novel Image Retrieval Based on a Combination of Local and Global Histograms of Visual Words},'' \emph{Mathematical Problems in Engineering}, vol. 2016, p. 8217250, 8 2016.

\bibitem{spitial4}
\BIBentryALTinterwordspacing
M.~Naeem, R.~Ashraf, N.~Ali, M.~Ahmad, and M.~A. Habib, ``{Bottom up Approach for Better Requirements Elicitation},'' in \emph{Proceedings of the International Conference on Future Networks and Distributed Systems}, ser. ICFNDS '17.\hskip 1em plus 0.5em minus 0.4em\relax New York, NY, USA: Association for Computing Machinery, 2017. [Online]. Available: \url{https://doi.org/10.1145/3102304.3109820}
\BIBentrySTDinterwordspacing

\bibitem{spitial5}
\BIBentryALTinterwordspacing
B.~Zafar, R.~Ashraf, N.~Ali, M.~Iqbal, M.~Sajid, S.~Dar, and N.~Ratyal, ``{A Novel Discriminating and Relative Global Spatial Image Representation with Applications in CBIR},'' \emph{Applied Sciences}, vol.~8, no.~11, p. 2242, Nov 2018. [Online]. Available: \url{http://dx.doi.org/10.3390/app8112242}
\BIBentrySTDinterwordspacing

\bibitem{spitial6}
H.~Anwar, S.~Zambanini, and M.~Kampel, ``{A rotation-invariant bag of visual words model for symbols based ancient coin classification},'' in \emph{2014 IEEE International Conference on Image Processing (ICIP)}, 2014, pp. 5257--5261.

\bibitem{spitial7}
\BIBentryALTinterwordspacing
R.~Khan, C.~Barat, D.~Muselet, and C.~Ducottet, ``{Spatial histograms of soft pairwise similar patches to improve the bag-of-visual-words model},'' \emph{Computer Vision and Image Understanding}, vol. 132, pp. 102--112, 2015. [Online]. Available: \url{https://www.sciencedirect.com/science/article/pii/S1077314214001878}
\BIBentrySTDinterwordspacing

\bibitem{fusion1}
R.~Ashraf, M.~Ahmed, U.~Ahmad, M.~A. Habib, S.~Jabbar, and K.~Naseer, ``{{MDCBIR-MF}: multimedia data for content-based image retrieval by using multiple features},'' \emph{Multimedia Tools and Applications}, vol.~79, no.~13, pp. 8553--8579, 4 2020.

\bibitem{fusion2}
\BIBentryALTinterwordspacing
Y.~Mistry, D.~Ingole, and M.~Ingole, ``{Content based image retrieval using hybrid features and various distance metric},'' \emph{Journal of Electrical Systems and Information Technology}, vol.~5, no.~3, pp. 874--888, 2018. [Online]. Available: \url{https://www.sciencedirect.com/science/article/pii/S2314717216301155}
\BIBentrySTDinterwordspacing

\bibitem{fusion3}
\BIBentryALTinterwordspacing
K.~T. Ahmed, S.~Ummesafi, and A.~Iqbal, ``{Content based image retrieval using image features information fusion},'' \emph{Information Fusion}, vol.~51, pp. 76--99, 2019. [Online]. Available: \url{https://www.sciencedirect.com/science/article/pii/S1566253517307893}
\BIBentrySTDinterwordspacing

\bibitem{fusion4}
\BIBentryALTinterwordspacing
P.~Liu, J.-M. Guo, K.~Chamnongthai, and H.~Prasetyo, ``{Fusion of color histogram and LBP-based features for texture image retrieval and classification},'' \emph{Information Sciences}, vol. 390, pp. 95--111, 2017. [Online]. Available: \url{https://www.sciencedirect.com/science/article/pii/S0020025517301159}
\BIBentrySTDinterwordspacing

\bibitem{fusion5}
A.~Nazir, R.~Ashraf, T.~Hamdani, and N.~Ali, ``{Content based image retrieval system by using HSV color histogram, discrete wavelet transform and edge histogram descriptor},'' in \emph{2018 International Conference on Computing, Mathematics and Engineering Technologies (iCoMET)}, 2018, pp. 1--6.

\bibitem{local1}
L.-W. Kang, C.-Y. Hsu, H.-W. Chen, C.-S. Lu, C.-Y. Lin, and S.-C. Pei, ``{Feature-Based Sparse Representation for Image Similarity Assessment},'' \emph{IEEE Transactions on Multimedia}, vol.~13, no.~5, pp. 1019--1030, 2011.

\bibitem{local2}
Z.-Q. Zhao, H.~Glotin, Z.~Xie, J.~Gao, and X.~Wu, ``{Cooperative Sparse Representation in Two Opposite Directions for Semi-Supervised Image Annotation},'' \emph{IEEE Transactions on Image Processing}, vol.~21, no.~9, pp. 4218--4231, 2012.

\bibitem{local3}
J.~J. Thiagarajan, K.~Natesan~Ramamurthy, P.~Sattigeri, and A.~Spanias, ``{Supervised local sparse coding of sub-image features for image retrieval},'' in \emph{2012 19th IEEE International Conference on Image Processing}, 2012, pp. 3117--3120.

\bibitem{local4}
\BIBentryALTinterwordspacing
D.~Wang, S.~C. Hoi, Y.~He, and J.~Zhu, ``{Retrieval-Based Face Annotation by Weak Label Regularized Local Coordinate Coding},'' in \emph{Proceedings of the 19th ACM International Conference on Multimedia}, ser. MM '11.\hskip 1em plus 0.5em minus 0.4em\relax New York, NY, USA: Association for Computing Machinery, 2011, p. 353–362. [Online]. Available: \url{https://doi.org/10.1145/2072298.2072345}
\BIBentrySTDinterwordspacing

\bibitem{local5}
\BIBentryALTinterwordspacing
C.~Hong and J.~Zhu, ``{Hypergraph-based multi-example ranking with sparse representation for transductive learning image retrieval},'' \emph{Neurocomputing}, vol. 101, pp. 94--103, 2013. [Online]. Available: \url{https://www.sciencedirect.com/science/article/pii/S0925231212006443}
\BIBentrySTDinterwordspacing

\bibitem{local6}
\BIBentryALTinterwordspacing
S.~Mohamadzadeh and H.~Farsi, ``{Content-based image retrieval system via sparse representation},'' \emph{IET Computer Vision}, vol.~10, no.~1, pp. 95--102, 2016. [Online]. Available: \url{https://ietresearch.onlinelibrary.wiley.com/doi/abs/10.1049/iet-cvi.2015.0165}
\BIBentrySTDinterwordspacing

\bibitem{local7}
Q.~Li, Y.~Han, and J.~Dang, ``{{Sketch4Image}: a novel framework for sketch-based image retrieval based on product quantization with coding residuals},'' \emph{Multimedia Tools and Applications}, vol.~75, no.~5, pp. 2419--2434, 3 2016.

\bibitem{local8}
Y.~Duan, J.~Lu, J.~Feng, and J.~Zhou, ``{Context-Aware Local Binary Feature Learning for Face Recognition},'' \emph{IEEE Transactions on Pattern Analysis and Machine Intelligence}, vol.~40, no.~5, pp. 1139--1153, 2018.

\bibitem{28_fesim2}
\BIBentryALTinterwordspacing
P.~Shamna, V.~Govindan, and K.~{Abdul Nazeer}, ``Content-based medical image retrieval by spatial matching of visual words,'' \emph{Journal of King Saud University - Computer and Information Sciences}, vol.~34, no.~2, pp. 58--71, 2022. [Online]. Available: \url{https://www.sciencedirect.com/science/article/pii/S131915781830750X}
\BIBentrySTDinterwordspacing

\bibitem{24_fabric01}
D.~Mo, W.~K. Wong, X.~Liu, and Y.~Ge, ``"concentrated hashing with neighborhood embedding for image retrieval and classification",'' \emph{International Journal of Machine Learning and Cybernetics}, vol.~13, no.~6, pp. 1571--1587, 6 2022.

\bibitem{25_med}
G.~Deep, J.~Kaur, S.~P. Singh, S.~R. Nayak, M.~Kumar, and S.~Kautish, ``"{MeQryEP}: A texture based descriptor for biomedical image retrieval",'' \emph{Journal of Healthcare Engineering}, vol. 2022, p. 9505229, 4 2022.

\bibitem{via}
\BIBentryALTinterwordspacing
A.~Dutta and A.~Zisserman, ``The {VIA} annotation software for images, audio and video,'' in \emph{Proceedings of the 27th ACM International Conference on Multimedia}, ser. MM '19.\hskip 1em plus 0.5em minus 0.4em\relax New York, NY, USA: ACM, 2019. [Online]. Available: \url{https://doi.org/10.1145/3343031.3350535}
\BIBentrySTDinterwordspacing

\bibitem{labelme}
B.~C. Russell, A.~Torralba, K.~P. Murphy, and W.~T. Freeman, ``{LabelMe}: A database and {Web-Based} tool for image annotation,'' \emph{International Journal of Computer Vision}, vol.~77, no.~1, pp. 157--173, 5 2008.

\bibitem{opencv}
G.~Bradski, ``{The OpenCV Library},'' \emph{Dr. Dobb's Journal of Software Tools}, 2000.

\bibitem{19_lakeicedata}
\BIBentryALTinterwordspacing
R.~Prabha, M.~Tom, M.~Rothermel, E.~Baltsavias, L.~Leal-Taixe, and K.~Schindler, ``{LAKE ICE MONITORING WITH WEBCAMS AND CROWD-SOURCED IMAGES},'' \emph{ISPRS Annals of the Photogrammetry, Remote Sensing and Spatial Information Sciences}, vol. V-2-2020, pp. 549--556, 2020. [Online]. Available: \url{https://isprs-annals.copernicus.org/articles/V-2-2020/549/2020/}
\BIBentrySTDinterwordspacing

\bibitem{17_chestCTdata}
A.~Ter-Sarkisov, ``{COVID-CT-Mask-Net: Prediction of COVID-19 from CT Scans Using Regional Features},'' \emph{Applied Intelligence}, vol.~52, p. 9664–9675, 2022.

\bibitem{18_heatsinkdata}
K.~Yang, Y.~Liu, S.~Zhang, and J.~Cao, ``{Surface Defect Detection of Heat Sink Based on Lightweight Fully Convolutional Network},'' \emph{IEEE Transactions on Instrumentation and Measurement}, vol.~71, pp. 1--12, 2022.

\bibitem{09_sift}
D.~G. Lowe, ``Distinctive image features from {Scale-Invariant} keypoints,'' \emph{International Journal of Computer Vision}, vol.~60, no.~2, pp. 91--110, 11 2004.

\bibitem{10_lbp}
T.~Ojala, M.~Pietikainen, and D.~Harwood, ``{Performance evaluation of texture measures with classification based on Kullback discrimination of distributions},'' in \emph{Proceedings of 12th International Conference on Pattern Recognition}, vol.~1, 1994, pp. 582--585 vol.1.

\end{thebibliography}






\newpage
\begin{appendices}
\section{Full ImR Evaluation Result for the Chest CT Dataset}
\label{app:chestct}
\begin{table}[H] 
\caption{Mean average precision and standard deviation values. The highest average precision with relative lowest standard deviation are marked in \textbf{bold text}.}
\label{tab:chestctm}
\footnotesize
\centering
\begin{tabular}{p{0.45in}p{0.5in}p{0.5in}cccccc}
\toprule
Feature & Similarity Metric & Image Size & $\mathrm{mAP}@1$ & $\mathrm{mAP}@5$ & $\mathrm{mAP}@10$ & $\mathrm{mAP}@15$ & $\mathrm{mAP}@20$ & Average\\\midrule
DefChars & Cosine & Raw & \textbf{0.88 $\pm$   0.06} & \textbf{0.86 $\pm$ 0.06} & \textbf{0.85 $\pm$ 0.06} & \textbf{0.84 $\pm$ 0.06} & \textbf{0.84 $\pm$ 0.06} & \textbf{0.85 $\pm$ 0.06} \\
DefChars & Euclidean & Raw & \textbf{0.88 $\pm$ 0.06} & \textbf{0.86 $\pm$ 0.06} & \textbf{0.85 $\pm$ 0.06} & \textbf{0.84 $\pm$ 0.06} & \textbf{0.84 $\pm$ 0.06} & \textbf{0.85 $\pm$ 0.06} \\
DefChars & Jaccard & Raw & 0.51 $\pm$ 0.19 & 0.49 $\pm$ 0.20 & 0.47 $\pm$ 0.21 & 0.47 $\pm$ 0.21 & 0.47 $\pm$ 0.22 & 0.48 $\pm$ 0.21 \\
DefChars & Manhattan & Raw & \textbf{0.88 $\pm$ 0.06} & 0.86 $\pm$ 0.07 & 0.85 $\pm$ 0.07 & 0.84 $\pm$ 0.07 & 0.84 $\pm$ 0.07 & 0.85 $\pm$ 0.07 \\ \midrule
Image & MSE & 8 & \textbf{0.76 $\pm$ 0.22} & \textbf{0.74 $\pm$ 0.22} & \textbf{0.73 $\pm$ 0.22} & \textbf{0.72 $\pm$ 0.22} & \textbf{0.71 $\pm$ 0.22} & \textbf{0.73 $\pm$ 0.22} \\
Image & MSE & 20 & 0.75 $\pm$ 0.24 & 0.74 $\pm$ 0.23 & 0.73 $\pm$ 0.23 & 0.72 $\pm$ 0.23 & 0.71 $\pm$ 0.23 & 0.73 $\pm$ 0.23 \\
Image & MSE & 50 & 0.75 $\pm$ 0.23 & 0.73 $\pm$ 0.23 & 0.73 $\pm$ 0.23 & 0.72 $\pm$ 0.23 & 0.71 $\pm$ 0.23 & 0.73 $\pm$ 0.23 \\
\bottomrule
\end{tabular}

\end{table}

\begin{table}[H] 
\caption{Average precision and standard deviation values of class 1. The highest average precision with relative lowest standard deviation are marked in \textbf{bold text}.}
\label{tab:chestct1}
\footnotesize
\centering
\begin{tabular}{p{0.45in}p{0.5in}p{0.5in}ccccc}
\toprule
Feature & Similarity Metric & Image Size & $\mathrm{AP}@1$ & $\mathrm{AP}@5$ & $\mathrm{AP}@10$ & $\mathrm{AP}@15$ & $\mathrm{AP}@20$ \\\midrule
DefChars & Cosine & Raw & 0.93 $\pm$   0.25 & \textbf{0.92 $\pm$ 0.21} & 0.91 $\pm$ 0.21 & 0.90 $\pm$ 0.21 & 0.90 $\pm$ 0.21 \\
DefChars & Euclidean & Raw & 0.93 $\pm$ 0.25 & \textbf{0.92 $\pm$ 0.21} & 0.91 $\pm$ 0.21 & 0.90 $\pm$ 0.21 & 0.90 $\pm$ 0.21 \\
DefChars & Jaccard & Raw & 0.70 $\pm$ 0.46 & 0.69 $\pm$ 0.33 & 0.68 $\pm$ 0.30 & 0.68 $\pm$ 0.29 & 0.68 $\pm$ 0.28 \\
DefChars & Manhattan & Raw & \textbf{0.94 $\pm$ 0.24} & \textbf{0.92 $\pm$ 0.21} & \textbf{0.92 $\pm$ 0.20} & \textbf{0.91 $\pm$ 0.20} & \textbf{0.91 $\pm$ 0.20} \\\midrule
Image & MSE & 8 & 0.93 $\pm$ 0.25 & 0.92 $\pm$ 0.20 & 0.91 $\pm$ 0.20 & 0.90 $\pm$ 0.20 & 0.89 $\pm$ 0.20 \\
Image & MSE & 20 & 0.94 $\pm$ 0.24 & 0.93 $\pm$ 0.19 & 0.92 $\pm$ 0.19 & 0.91 $\pm$ 0.19 & \textbf{0.91 $\pm$ 0.19} \\
Image & MSE & 50 & \textbf{0.94 $\pm$ 0.23} & \textbf{0.93 $\pm$ 0.18} & \textbf{0.92 $\pm$ 0.18} & \textbf{0.92 $\pm$ 0.19} & \textbf{0.91 $\pm$ 0.19} \\
\bottomrule
\end{tabular}

\end{table}

\begin{table}[H] 
\caption{Average precision and standard deviation values of class 2. The highest average precision with relative lowest standard deviation are marked in \textbf{bold text}.}
\label{tab:chestct2}
\footnotesize
\centering
\begin{tabular}{p{0.45in}p{0.5in}p{0.5in}ccccc}
\toprule
Feature & Similarity Metric & Image Size & $\mathrm{AP}@1$ & $\mathrm{AP}@5$ & $\mathrm{AP}@10$ & $\mathrm{AP}@15$ & $\mathrm{AP}@20$ \\\midrule
DefChars & Cosine & Raw & 0.88 $\pm$   0.32 & 0.86 $\pm$ 0.24 & 0.85 $\pm$ 0.23 & 0.84 $\pm$ 0.23 & 0.83 $\pm$ 0.23 \\
DefChars & Euclidean & Raw & \textbf{0.89 $\pm$ 0.31} & 0.86 $\pm$ 0.24 & \textbf{0.85 $\pm$ 0.22} & 0.84 $\pm$ 0.22 & 0.84 $\pm$ 0.22 \\
DefChars & Jaccard & Raw & 0.50 $\pm$ 0.50 & 0.48 $\pm$ 0.25 & 0.47 $\pm$ 0.19 & 0.47 $\pm$ 0.17 & 0.47 $\pm$ 0.15 \\
DefChars & Manhattan & Raw & 0.88 $\pm$ 0.33 & \textbf{0.86 $\pm$ 0.23} & \textbf{0.85 $\pm$ 0.22} & \textbf{0.85 $\pm$ 0.22} & \textbf{0.84 $\pm$ 0.21} \\\midrule
Image & MSE & 8 & \textbf{0.82 $\pm$ 0.38} & \textbf{0.80 $\pm$ 0.28} & \textbf{0.79 $\pm$ 0.26} & \textbf{0.78 $\pm$ 0.25} & \textbf{0.78 $\pm$ 0.25} \\
Image & MSE & 20 & 0.82 $\pm$ 0.39 & 0.79 $\pm$ 0.28 & 0.78 $\pm$ 0.26 & \textbf{0.78 $\pm$ 0.25} & 0.77 $\pm$ 0.25 \\
Image & MSE & 50 & 0.81 $\pm$ 0.39 & 0.79 $\pm$ 0.29 & 0.78 $\pm$ 0.26 & \textbf{0.78 $\pm$ 0.25} & 0.77 $\pm$ 0.25 \\
\bottomrule
\end{tabular}

\end{table}

\begin{table}[H] 
\caption{Average precision and standard deviation values of class 3. The highest average precision with relative lowest standard deviation are marked in \textbf{bold text}.}
\label{tab:chestct3}
\footnotesize
\centering
\begin{tabular}{p{0.45in}p{0.5in}p{0.5in}ccccc}
\toprule
Feature & Similarity Metric & Image Size & $\mathrm{AP}@1$ & $\mathrm{AP}@5$ & $\mathrm{AP}@10$ & $\mathrm{AP}@15$ & $\mathrm{AP}@20$ \\\midrule
DefChars & Cosine & Raw & 0.82 $\pm$   0.39 & \textbf{0.81 $\pm$ 0.26} & \textbf{0.80 $\pm$ 0.24} & \textbf{0.78 $\pm$ 0.23} & \textbf{0.77 $\pm$ 0.22} \\
DefChars & Euclidean & Raw & 0.81 $\pm$ 0.39 & 0.81 $\pm$ 0.27 & 0.79 $\pm$ 0.24 & \textbf{0.78 $\pm$ 0.23} & 0.77 $\pm$ 0.23 \\
DefChars & Jaccard & Raw & 0.33 $\pm$ 0.47 & 0.29 $\pm$ 0.21 & 0.26 $\pm$ 0.15 & 0.25 $\pm$ 0.12 & 0.25 $\pm$ 0.11 \\
DefChars & Manhattan & Raw & \textbf{0.82 $\pm$ 0.38} & 0.79 $\pm$ 0.27 & 0.78 $\pm$ 0.24 & 0.77 $\pm$ 0.23 & \textbf{0.77 $\pm$ 0.22} \\\midrule
Image & MSE & 8 & \textbf{0.52 $\pm$ 0.50} & \textbf{0.50 $\pm$ 0.32} & \textbf{0.48 $\pm$ 0.28} & \textbf{0.47 $\pm$ 0.26} & \textbf{0.46 $\pm$ 0.25} \\
Image & MSE & 20 & 0.48 $\pm$ 0.50 & 0.48 $\pm$ 0.32 & 0.48 $\pm$ 0.29 & 0.47 $\pm$ 0.27 & 0.46 $\pm$ 0.26 \\
Image & MSE & 50 & 0.49 $\pm$ 0.50 & 0.48 $\pm$ 0.32 & 0.47 $\pm$ 0.29 & 0.46 $\pm$ 0.27 & 0.46 $\pm$ 0.26 \\
\bottomrule
\end{tabular}

\end{table}

\section{Full ImR Evaluation Result for the Heatsink Dataset}
\label{app:heatsink}

\begin{table}[H] 
\caption{Mean average precision and standard deviation values. The highest average precision with relative lowest standard deviation are marked in \textbf{bold text}.}
\label{tab:heatsinkm}
\footnotesize
\centering
\begin{tabular}{p{0.45in}p{0.5in}p{0.5in}cccccc}
\toprule
Feature & Similarity Metric & Image Size & $\mathrm{mAP}@1$ & $\mathrm{mAP}@5$ & $\mathrm{mAP}@10$ & $\mathrm{mAP}@15$ & $\mathrm{mAP}@20$ & Average \\\midrule
DefChars & Cosine & Raw & \textbf{0.98 $\pm$   0.01} & \textbf{0.97 $\pm$ 0.02} & \textbf{0.97 $\pm$ 0.02} & \textbf{0.97 $\pm$ 0.03} & \textbf{0.96 $\pm$ 0.03} & \textbf{0.97 $\pm$ 0.02} \\
DefChars & Euclidean & Raw & \textbf{0.98 $\pm$ 0.01} & \textbf{0.97 $\pm$ 0.02} & \textbf{0.97 $\pm$ 0.02} & \textbf{0.97 $\pm$ 0.03} & \textbf{0.96 $\pm$ 0.03} & \textbf{0.97 $\pm$ 0.02} \\
DefChars & Jaccard & Raw & 0.64 $\pm$ 0.41 & 0.63 $\pm$ 0.43 & 0.62 $\pm$ 0.43 & 0.61 $\pm$ 0.44 & 0.61 $\pm$ 0.44 & 0.62 $\pm$ 0.43 \\
DefChars & Manhattan & Raw & \textbf{0.98 $\pm$ 0.01} & \textbf{0.97 $\pm$ 0.02} & 0.97 $\pm$ 0.03 & 0.96 $\pm$ 0.03 & \textbf{0.96 $\pm$ 0.03} & \textbf{0.97 $\pm$ 0.02} \\ \midrule
Image & MSE & 8 & \textbf{0.89 $\pm$ 0.11} & \textbf{0.88 $\pm$ 0.13} & \textbf{0.87 $\pm$ 0.13} & \textbf{0.87 $\pm$ 0.14} & \textbf{0.86 $\pm$ 0.14} & \textbf{0.87 $\pm$ 0.13} \\
Image & MSE & 20 & 0.89 $\pm$ 0.13 & 0.87 $\pm$ 0.14 & 0.87 $\pm$ 0.14 & 0.86 $\pm$ 0.15 & 0.86 $\pm$ 0.15 & 0.87 $\pm$ 0.14 \\
Image & MSE & 50 & 0.88 $\pm$ 0.13 & 0.87 $\pm$ 0.15 & 0.86 $\pm$ 0.15 & 0.86 $\pm$ 0.15 & 0.86 $\pm$ 0.15 & 0.87 $\pm$ 0.15 \\
\bottomrule
\end{tabular}

\end{table}

\begin{table}[H] 
\caption{Average precision and standard deviation values of class 1. The highest average precision with relative lowest standard deviation are marked in \textbf{bold text}.}
\label{tab:heatsink1}
\footnotesize
\centering
\begin{tabular}{p{0.45in}p{0.5in}p{0.5in}ccccc}
\toprule
Feature & Similarity Metric & Image Size & $\mathrm{AP}@1$ & $\mathrm{AP}@5$ & $\mathrm{AP}@10$ & $\mathrm{AP}@15$ & $\mathrm{AP}@20$ \\\midrule
DefChars & Cosine & Raw & \textbf{0.98 $\pm$   0.15} & \textbf{0.96 $\pm$ 0.15} & \textbf{0.95 $\pm$ 0.15} & \textbf{0.95 $\pm$ 0.16} & \textbf{0.94 $\pm$ 0.16} \\
DefChars & Euclidean & Raw & \textbf{0.98 $\pm$ 0.15} & \textbf{0.96 $\pm$ 0.15} & \textbf{0.95 $\pm$ 0.15} & \textbf{0.95 $\pm$ 0.16} & \textbf{0.94 $\pm$ 0.16} \\
DefChars & Jaccard & Raw & 0.35 $\pm$ 0.48 & 0.32 $\pm$ 0.26 & 0.31 $\pm$ 0.21 & 0.31 $\pm$ 0.18 & 0.30 $\pm$ 0.17 \\
DefChars & Manhattan & Raw & 0.97 $\pm$ 0.17 & 0.96 $\pm$ 0.16 & 0.95 $\pm$ 0.16 & 0.94 $\pm$ 0.16 & 0.94 $\pm$ 0.17 \\\midrule
Image & MSE & 8 & \textbf{0.81 $\pm$ 0.39} & \textbf{0.79 $\pm$ 0.31} & \textbf{0.77 $\pm$ 0.30} & \textbf{0.77 $\pm$ 0.29} & \textbf{0.76 $\pm$ 0.29} \\
Image & MSE & 20 & 0.79 $\pm$ 0.40 & 0.77 $\pm$ 0.31 & 0.76 $\pm$ 0.30 & 0.76 $\pm$ 0.30 & 0.75 $\pm$ 0.29 \\
Image & MSE & 50 & 0.79 $\pm$ 0.41 & 0.77 $\pm$ 0.32 & 0.76 $\pm$ 0.30 & 0.75 $\pm$ 0.30 & 0.75 $\pm$ 0.30 \\
\bottomrule
\end{tabular}

\end{table}

\begin{table}[H] 
\caption{Average precision and standard deviation values of class 2. The highest average precision with relative lowest standard deviation are marked in \textbf{bold text}.}
\label{tab:heatsink2}
\footnotesize
\centering
\begin{tabular}{p{0.45in}p{0.5in}p{0.5in}ccccc}
\toprule
Feature & Similarity Metric & Image Size & $\mathrm{AP}@1$ & $\mathrm{AP}@5$ & $\mathrm{AP}@10$ & $\mathrm{AP}@15$ & $\mathrm{AP}@20$ \\\midrule
DefChars & Cosine & Raw & \textbf{0.99 $\pm$   0.10} & \textbf{0.99 $\pm$ 0.07} & \textbf{0.99 $\pm$ 0.07} & \textbf{0.98 $\pm$ 0.08} & \textbf{0.98 $\pm$ 0.08} \\
DefChars & Euclidean & Raw & \textbf{0.99 $\pm$ 0.10} & \textbf{0.99 $\pm$ 0.07} & \textbf{0.99 $\pm$ 0.07} & \textbf{0.98 $\pm$ 0.08} & \textbf{0.98 $\pm$ 0.08} \\
DefChars & Jaccard & Raw & 0.94 $\pm$ 0.25 & 0.93 $\pm$ 0.13 & 0.92 $\pm$ 0.10 & 0.92 $\pm$ 0.09 & 0.92 $\pm$ 0.09 \\
DefChars & Manhattan & Raw & \textbf{0.99 $\pm$ 0.10} & \textbf{0.99 $\pm$ 0.07} & 0.98 $\pm$ 0.08 & \textbf{0.98 $\pm$ 0.08} & \textbf{0.98 $\pm$ 0.08} \\\midrule
Image & MSE & 8 & 0.97 $\pm$ 0.17 & 0.97 $\pm$ 0.13 & 0.96 $\pm$ 0.13 & 0.96 $\pm$ 0.13 & 0.96 $\pm$ 0.13 \\
Image & MSE & 20 & \textbf{0.98 $\pm$ 0.15} & \textbf{0.97 $\pm$ 0.12} & 0.97 $\pm$ 0.13 & \textbf{0.97 $\pm$ 0.13} & \textbf{0.97 $\pm$ 0.13} \\
Image & MSE & 50 & \textbf{0.98 $\pm$ 0.15} & \textbf{0.97 $\pm$ 0.12} & \textbf{0.97 $\pm$ 0.12} & \textbf{0.97 $\pm$ 0.13} & \textbf{0.97 $\pm$ 0.13} \\
\bottomrule
\end{tabular}

\end{table}

\section{Full ImR Evaluation Result for the Lake Ice Dataset}
\label{app:lakeice}
\begin{table}[H] 
\caption{Mean average precision and standard deviation values. The highest average precision with relative lowest standard deviation are marked in \textbf{bold text}.}
\label{tab:lakeicem}
\footnotesize
\centering
\begin{tabular}{p{0.45in}p{0.5in}p{0.5in}cccccc}
\toprule
Feature & Similarity Metric & Image Size & $\mathrm{mAP}@1$ & $\mathrm{mAP}@5$ & $\mathrm{mAP}@10$ & $\mathrm{mAP}@15$ & $\mathrm{mAP}@20$ & Average \\\midrule
DefChars & Cosine & Raw & 0.95 $\pm$   0.03 & 0.90 $\pm$ 0.06 & 0.87 $\pm$ 0.09 & 0.85 $\pm$ 0.10 & 0.84 $\pm$ 0.12 & 0.88 $\pm$ 0.08 \\
DefChars & Euclidean & Raw & 0.95 $\pm$ 0.03 & 0.90 $\pm$ 0.07 & 0.87 $\pm$ 0.09 & 0.85 $\pm$ 0.11 & 0.83 $\pm$ 0.13 & 0.88 $\pm$ 0.09 \\
DefChars & Jaccard & Raw & 0.84 $\pm$ 0.12 & 0.80 $\pm$ 0.17 & 0.77 $\pm$ 0.20 & 0.74 $\pm$ 0.24 & 0.72 $\pm$ 0.27 & 0.77 $\pm$ 0.20 \\
DefChars & Manhattan & Raw & \textbf{0.96 $\pm$ 0.03} & \textbf{0.92 $\pm$ 0.05} & \textbf{0.89 $\pm$ 0.07} & \textbf{0.87 $\pm$ 0.09} & \textbf{0.86 $\pm$ 0.11} & \textbf{0.90 $\pm$ 0.07} \\ \midrule
Image & MSE & 8 & \textbf{0.93 $\pm$ 0.08} & \textbf{0.87 $\pm$ 0.14} & \textbf{0.82 $\pm$ 0.17} & \textbf{0.80 $\pm$ 0.19} & \textbf{0.78 $\pm$ 0.21} & \textbf{0.84 $\pm$ 0.16} \\
Image & MSE & 20 & 0.93 $\pm$ 0.09 & 0.86 $\pm$ 0.16 & 0.82 $\pm$ 0.19 & 0.79 $\pm$ 0.21 & 0.78 $\pm$ 0.23 & 0.84 $\pm$ 0.18 \\
Image & MSE & 50 & 0.93 $\pm$ 0.10 & 0.86 $\pm$ 0.16 & 0.82 $\pm$ 0.19 & 0.79 $\pm$ 0.21 & 0.77 $\pm$ 0.23 & 0.83 $\pm$ 0.18 \\
\bottomrule
\end{tabular}

\end{table}

\begin{table}[H] 
\caption{Average precision and standard deviation values of class 1. The highest average precision with relative lowest standard deviation are marked in \textbf{bold text}.}
\label{tab:lakeice1}
\footnotesize
\centering
\begin{tabular}{p{0.45in}p{0.5in}p{0.5in}ccccc}
\toprule
Feature & Similarity Metric & Image Size & $\mathrm{AP}@1$ & $\mathrm{AP}@5$ & $\mathrm{AP}@10$ & $\mathrm{AP}@15$ & $\mathrm{AP}@20$ \\\midrule
DefChars & Cosine & Raw & 0.92 $\pm$   0.28 & 0.87 $\pm$ 0.26 & 0.84 $\pm$ 0.27 & 0.83 $\pm$ 0.27 & 0.83 $\pm$ 0.28 \\
DefChars & Euclidean & Raw & 0.92 $\pm$ 0.27 & 0.86 $\pm$ 0.26 & 0.84 $\pm$ 0.27 & 0.83 $\pm$ 0.27 & 0.83 $\pm$ 0.27 \\
DefChars & Jaccard & Raw & 0.80 $\pm$ 0.40 & 0.77 $\pm$ 0.27 & 0.75 $\pm$ 0.26 & 0.72 $\pm$ 0.25 & 0.72 $\pm$ 0.24 \\
DefChars & Manhattan & Raw & \textbf{0.94 $\pm$ 0.24} & \textbf{0.88 $\pm$ 0.24} & \textbf{0.86 $\pm$ 0.25} & \textbf{0.85 $\pm$ 0.26} & \textbf{0.84 $\pm$ 0.26} \\\midrule
Image & MSE & 8 & 0.94 $\pm$ 0.23 & 0.89 $\pm$ 0.24 & 0.87 $\pm$ 0.25 & 0.86 $\pm$ 0.26 & 0.85 $\pm$ 0.25 \\
Image & MSE & 20 & \textbf{0.97 $\pm$ 0.18} & \textbf{0.91 $\pm$ 0.22} & 0.88 $\pm$ 0.25 & \textbf{0.87 $\pm$ 0.25} & \textbf{0.86 $\pm$ 0.25} \\
Image & MSE & 50 & \textbf{0.97 $\pm$ 0.18} & \textbf{0.91 $\pm$ 0.22} & \textbf{0.88 $\pm$ 0.24} & \textbf{0.87 $\pm$ 0.25} & \textbf{0.86 $\pm$ 0.25} \\
\bottomrule
\end{tabular}

\end{table}

\begin{table}[H] 
\caption{Average precision and standard deviation values of class 2. The highest average precision with relative lowest standard deviation are marked in \textbf{bold text}.}
\label{tab:lakeice2}
\footnotesize
\centering
\begin{tabular}{p{0.45in}p{0.5in}p{0.5in}ccccc}
\toprule
Feature & Similarity Metric & Image Size & $\mathrm{AP}@1$ & $\mathrm{AP}@5$ & $\mathrm{AP}@10$ & $\mathrm{AP}@15$ & $\mathrm{AP}@20$ \\\midrule
DefChars & Cosine & Raw & 0.97 $\pm$   0.17 & 0.93 $\pm$ 0.20 & 0.91 $\pm$ 0.22 & 0.89 $\pm$ 0.23 & 0.88 $\pm$ 0.24 \\
DefChars & Euclidean & Raw & 0.97 $\pm$ 0.17 & 0.93 $\pm$ 0.19 & 0.91 $\pm$ 0.22 & 0.90 $\pm$ 0.23 & 0.89 $\pm$ 0.24 \\
DefChars & Jaccard & Raw & 0.91 $\pm$ 0.28 & 0.91 $\pm$ 0.26 & 0.90 $\pm$ 0.27 & 0.90 $\pm$ 0.27 & \textbf{0.90 $\pm$ 0.28} \\
DefChars & Manhattan & Raw & \textbf{0.98 $\pm$ 0.14} & \textbf{0.95 $\pm$ 0.17} & \textbf{0.92 $\pm$ 0.20} & \textbf{0.91 $\pm$ 0.21} & \textbf{0.90 $\pm$ 0.22} \\\midrule
Image & MSE & 8 & 0.97 $\pm$ 0.18 & 0.92 $\pm$ 0.22 & 0.87 $\pm$ 0.25 & 0.83 $\pm$ 0.28 & 0.81 $\pm$ 0.29 \\
Image & MSE & 20 & \textbf{0.98 $\pm$ 0.14} & \textbf{0.93 $\pm$ 0.20} & \textbf{0.88 $\pm$ 0.24} & 0.84 $\pm$ 0.27 & 0.82 $\pm$ 0.29 \\
Image & MSE & 50 & 0.98 $\pm$ 0.15 & \textbf{0.93 $\pm$ 0.20} & \textbf{0.88 $\pm$ 0.24} & 0.84 $\pm$ 0.27 & 0.82 $\pm$ 0.29 \\
\bottomrule
\end{tabular}

\end{table}

\begin{table}[H] 
\caption{Average precision and standard deviation values of class 3. The highest average precision with relative lowest standard deviation are marked in \textbf{bold text}.}
\label{tab:lakeice3}
\footnotesize
\centering
\begin{tabular}{p{0.45in}p{0.5in}p{0.5in}ccccc}
\toprule
Feature & Similarity Metric & Image Size & $\mathrm{AP}@1$ & $\mathrm{AP}@5$ & $\mathrm{AP}@10$ & $\mathrm{AP}@15$ & $\mathrm{AP}@20$ \\\midrule
DefChars & Cosine & Raw & 0.99 $\pm$   0.12 & 0.98 $\pm$ 0.11 & 0.97 $\pm$ 0.12 & 0.97 $\pm$ 0.12 & 0.96 $\pm$ 0.13 \\
DefChars & Euclidean & Raw & 0.99 $\pm$ 0.12 & 0.98 $\pm$ 0.11 & 0.97 $\pm$ 0.12 & 0.96 $\pm$ 0.13 & 0.96 $\pm$ 0.13 \\
DefChars & Jaccard & Raw & 0.97 $\pm$ 0.18 & 0.95 $\pm$ 0.16 & 0.94 $\pm$ 0.16 & 0.93 $\pm$ 0.16 & 0.92 $\pm$ 0.16 \\
DefChars & Manhattan & Raw & \textbf{0.99 $\pm$ 0.10} & \textbf{0.98 $\pm$ 0.09} & \textbf{0.98 $\pm$ 0.10} & \textbf{0.97 $\pm$ 0.11} & \textbf{0.97 $\pm$ 0.11} \\\midrule
Image & MSE & 8 & 0.99 $\pm$ 0.10 & 0.98 $\pm$ 0.09 & 0.98 $\pm$ 0.10 & \textbf{0.98 $\pm$ 0.10} & 0.97 $\pm$ 0.10 \\
Image & MSE & 20 & 0.99 $\pm$ 0.10 & \textbf{0.99 $\pm$ 0.09} & \textbf{0.98 $\pm$ 0.09} & \textbf{0.98 $\pm$ 0.10} & \textbf{0.98 $\pm$ 0.10} \\
Image & MSE & 50 & \textbf{0.99 $\pm$ 0.09} & 0.98 $\pm$ 0.09 & \textbf{0.98 $\pm$ 0.09} & \textbf{0.98 $\pm$ 0.10} & \textbf{0.98 $\pm$ 0.10} \\
\bottomrule
\end{tabular}

\end{table}

\begin{table}[H] 
\caption{Average precision and standard deviation values of class 4. The highest average precision with relative lowest standard deviation are marked in \textbf{bold text}.}
\label{tab:lakeice4}
\footnotesize
\centering
\begin{tabular}{p{0.45in}p{0.5in}p{0.5in}ccccc}
\toprule
Feature & Similarity Metric & Image Size & $\mathrm{AP}@1$ & $\mathrm{AP}@5$ & $\mathrm{AP}@10$ & $\mathrm{AP}@15$ & $\mathrm{AP}@20$ \\\midrule
DefChars & Cosine & Raw & 0.93 $\pm$   0.25 & 0.85 $\pm$ 0.26 & 0.77 $\pm$ 0.28 & 0.72 $\pm$ 0.29 & 0.68 $\pm$ 0.29 \\
DefChars & Euclidean & Raw & 0.92 $\pm$ 0.27 & 0.83 $\pm$ 0.27 & 0.76 $\pm$ 0.28 & 0.71 $\pm$ 0.29 & 0.66 $\pm$ 0.30 \\
DefChars & Jaccard & Raw & 0.69 $\pm$ 0.46 & 0.58 $\pm$ 0.40 & 0.50 $\pm$ 0.38 & 0.40 $\pm$ 0.29 & 0.33 $\pm$ 0.23 \\
DefChars & Manhattan & Raw & \textbf{0.94 $\pm$ 0.23} & \textbf{0.89 $\pm$ 0.23} & \textbf{0.81 $\pm$ 0.25} & \textbf{0.76 $\pm$ 0.27} & \textbf{0.72 $\pm$ 0.27} \\\midrule
Image & MSE & 8 & \textbf{0.81 $\pm$ 0.40} & \textbf{0.67 $\pm$ 0.36} & \textbf{0.58 $\pm$ 0.35} & \textbf{0.52 $\pm$ 0.33} & \textbf{0.47 $\pm$ 0.31} \\
Image & MSE & 20 & 0.79 $\pm$ 0.41 & 0.63 $\pm$ 0.39 & 0.54 $\pm$ 0.37 & 0.49 $\pm$ 0.34 & 0.45 $\pm$ 0.32 \\
Image & MSE & 50 & 0.78 $\pm$ 0.41 & 0.63 $\pm$ 0.38 & 0.54 $\pm$ 0.37 & 0.49 $\pm$ 0.35 & 0.44 $\pm$ 0.33 \\
\bottomrule
\end{tabular}

\end{table}

\section{Full ImR Evaluation Result for the Wind Turbine Blade Dataset}
\label{app:wtb}

\begin{table}[H] 
\caption{Mean average precision and standard deviation values. The highest average precision with relative lowest standard deviation are marked in \textbf{bold text}.}
\label{tab:wtbm}
\footnotesize
\centering
\begin{tabular}{p{0.45in}p{0.5in}p{0.5in}cccccc}
\toprule
Feature & Similarity Metric & Image Size & $\mathrm{mAP}@1$ & $\mathrm{mAP}@5$ & $\mathrm{mAP}@10$ & $\mathrm{mAP}@15$ & $\mathrm{mAP}@20$ & Average \\\midrule
DefChars & Cosine & Raw & 0.74 $\pm$   0.06 & 0.62 $\pm$ 0.14 & 0.57 $\pm$ 0.16 & 0.54 $\pm$ 0.18 & 0.52 $\pm$ 0.19 & 0.60 $\pm$ 0.15 \\
DefChars & Euclidean & Raw & 0.75 $\pm$ 0.08 & 0.63 $\pm$ 0.15 & 0.58 $\pm$ 0.17 & 0.54 $\pm$ 0.19 & 0.52 $\pm$ 0.20 & 0.60 $\pm$ 0.16 \\
DefChars & Jaccard & Raw & 0.41 $\pm$ 0.20 & 0.41 $\pm$ 0.17 & 0.37 $\pm$ 0.18 & 0.36 $\pm$ 0.17 & 0.36 $\pm$ 0.17 & 0.38 $\pm$ 0.18 \\
DefChars & Manhattan & Raw & \textbf{0.77 $\pm$ 0.09} & \textbf{0.64 $\pm$ 0.17} & \textbf{0.59 $\pm$ 0.18} & \textbf{0.55 $\pm$ 0.20} & \textbf{0.53 $\pm$ 0.20} & \textbf{0.62 $\pm$ 0.17} \\ \midrule
Image & MSE & 8 & 0.52 $\pm$ 0.23 & \textbf{0.44 $\pm$ 0.33} & 0.42 $\pm$ 0.34 & \textbf{0.42 $\pm$ 0.34} & \textbf{0.40 $\pm$ 0.33} & \textbf{0.44 $\pm$ 0.31} \\
Image & MSE & 20 & 0.54 $\pm$ 0.28 & 0.44 $\pm$ 0.35 & \textbf{0.43 $\pm$ 0.36} & 0.41 $\pm$ 0.35 & 0.40 $\pm$ 0.35 & 0.44 $\pm$ 0.34 \\
Image & MSE & 50 & \textbf{0.54 $\pm$ 0.27} & 0.44 $\pm$ 0.36 & \textbf{0.43 $\pm$ 0.36} & 0.41 $\pm$ 0.35 & 0.40 $\pm$ 0.35 & 0.44 $\pm$ 0.34 \\
\bottomrule
\end{tabular}

\end{table}

\begin{table}[H] 
\caption{Average precision and standard deviation values of class 1. The highest average precision with relative lowest standard deviation are marked in \textbf{bold text}.}
\label{tab:wtb1}
\footnotesize
\centering
\begin{tabular}{p{0.45in}p{0.5in}p{0.5in}ccccc}
\toprule
Feature & Similarity Metric & Image Size & $\mathrm{AP}@1$ & $\mathrm{AP}@5$ & $\mathrm{AP}@10$ & $\mathrm{AP}@15$ & $\mathrm{AP}@20$ \\\midrule
DefChars & Cosine & Raw & 0.75 $\pm$   0.43 & 0.64 $\pm$ 0.30 & 0.59 $\pm$ 0.28 & \textbf{0.55 $\pm$ 0.26} & 0.52 $\pm$ 0.25 \\
DefChars & Euclidean & Raw & 0.73 $\pm$ 0.45 & 0.64 $\pm$ 0.30 & \textbf{0.59 $\pm$ 0.26} & \textbf{0.55 $\pm$ 0.26} & \textbf{0.53 $\pm$ 0.24} \\
DefChars & Jaccard & Raw & 0.31 $\pm$ 0.47 & 0.37 $\pm$ 0.23 & 0.31 $\pm$ 0.16 & 0.30 $\pm$ 0.14 & 0.29 $\pm$ 0.12 \\
DefChars & Manhattan & Raw & \textbf{0.78 $\pm$ 0.42} & \textbf{0.64 $\pm$ 0.29} & 0.59 $\pm$ 0.27 & \textbf{0.55 $\pm$ 0.26} & \textbf{0.53 $\pm$ 0.24} \\\midrule
Image & MSE & 8 & \textbf{0.71 $\pm$ 0.46} & 0.70 $\pm$ 0.36 & 0.68 $\pm$ 0.31 & 0.67 $\pm$ 0.29 & 0.65 $\pm$ 0.28 \\
Image & MSE & 20 & \textbf{0.71 $\pm$ 0.46} & 0.71 $\pm$ 0.36 & \textbf{0.73 $\pm$ 0.30} & 0.70 $\pm$ 0.30 & 0.69 $\pm$ 0.28 \\
Image & MSE & 50 & 0.70 $\pm$ 0.46 & 0.72 $\pm$ 0.36 & \textbf{0.73 $\pm$ 0.30} & 0.71 $\pm$ 0.29 & 0.69 $\pm$ 0.28 \\
\bottomrule
\end{tabular}

\end{table}

\begin{table}[H] 
\caption{Average precision and standard deviation values of class 2. The highest average precision with relative lowest standard deviation are marked in \textbf{bold text}.}
\label{tab:wtb2}
\footnotesize
\centering
\begin{tabular}{p{0.45in}p{0.5in}p{0.5in}ccccc}
\toprule
Feature & Similarity Metric & Image Size & $\mathrm{AP}@1$ & $\mathrm{AP}@5$ & $\mathrm{AP}@10$ & $\mathrm{AP}@15$ & $\mathrm{AP}@20$ \\\midrule
DefChars & Cosine & Raw & 0.68 $\pm$   0.47 & 0.65 $\pm$ 0.32 & 0.62 $\pm$ 0.27 & \textbf{0.62 $\pm$ 0.25} & \textbf{0.61 $\pm$ 0.24} \\
DefChars & Euclidean & Raw & 0.68 $\pm$ 0.47 & 0.64 $\pm$ 0.32 & 0.62 $\pm$ 0.28 & 0.59 $\pm$ 0.27 & 0.58 $\pm$ 0.26 \\
DefChars & Jaccard & Raw & 0.45 $\pm$ 0.50 & 0.42 $\pm$ 0.28 & 0.41 $\pm$ 0.20 & 0.41 $\pm$ 0.18 & 0.41 $\pm$ 0.17 \\
DefChars & Manhattan & Raw & \textbf{0.70 $\pm$ 0.46} & \textbf{0.66 $\pm$ 0.30} & \textbf{0.63 $\pm$ 0.29} & 0.60 $\pm$ 0.28 & 0.59 $\pm$ 0.27 \\\midrule
Image & MSE & 8 & 0.41 $\pm$ 0.50 & \textbf{0.25 $\pm$ 0.21} & \textbf{0.23 $\pm$ 0.17} & \textbf{0.21 $\pm$ 0.13} & \textbf{0.20 $\pm$ 0.12} \\
Image & MSE & 20 & 0.42 $\pm$ 0.50 & 0.23 $\pm$ 0.20 & 0.20 $\pm$ 0.16 & 0.19 $\pm$ 0.14 & 0.17 $\pm$ 0.12 \\
Image & MSE & 50 & \textbf{0.44 $\pm$ 0.50} & 0.22 $\pm$ 0.20 & 0.20 $\pm$ 0.15 & 0.18 $\pm$ 0.13 & 0.17 $\pm$ 0.11 \\
\bottomrule
\end{tabular}

\end{table}

\begin{table}[H] 
\caption{Average precision and standard deviation values of class 3. The highest average precision with relative lowest standard deviation are marked in \textbf{bold text}.}
\label{tab:wtb3}
\footnotesize
\centering
\begin{tabular}{p{0.45in}p{0.5in}p{0.5in}ccccc}
\toprule
Feature & Similarity Metric & Image Size & $\mathrm{AP}@1$ & $\mathrm{AP}@5$ & $\mathrm{AP}@10$ & $\mathrm{AP}@15$ & $\mathrm{AP}@20$ \\\midrule
DefChars & Cosine & Raw & 0.82 $\pm$   0.38 & 0.77 $\pm$ 0.27 & 0.73 $\pm$ 0.27 & 0.71 $\pm$ 0.27 & 0.69 $\pm$ 0.26 \\
DefChars & Euclidean & Raw & 0.87 $\pm$ 0.33 & 0.79 $\pm$ 0.26 & 0.75 $\pm$ 0.27 & 0.74 $\pm$ 0.26 & 0.72 $\pm$ 0.24 \\
DefChars & Jaccard & Raw & 0.68 $\pm$ 0.47 & 0.64 $\pm$ 0.30 & 0.60 $\pm$ 0.23 & 0.58 $\pm$ 0.20 & 0.56 $\pm$ 0.19 \\
DefChars & Manhattan & Raw & \textbf{0.90 $\pm$ 0.30} & \textbf{0.83 $\pm$ 0.25} & \textbf{0.78 $\pm$ 0.26} & \textbf{0.76 $\pm$ 0.25} & \textbf{0.74 $\pm$ 0.25} \\\midrule
Image & MSE & 8 & 0.70 $\pm$ 0.46 & 0.74 $\pm$ 0.29 & \textbf{0.74 $\pm$ 0.27} & \textbf{0.73 $\pm$ 0.25} & \textbf{0.71 $\pm$ 0.25} \\
Image & MSE & 20 & \textbf{0.82 $\pm$ 0.38} & \textbf{0.76 $\pm$ 0.29} & 0.74 $\pm$ 0.28 & 0.72 $\pm$ 0.26 & \textbf{0.71 $\pm$ 0.25} \\
Image & MSE & 50 & \textbf{0.82 $\pm$ 0.38} & \textbf{0.76 $\pm$ 0.29} & 0.74 $\pm$ 0.28 & 0.72 $\pm$ 0.27 & 0.71 $\pm$ 0.26 \\
\bottomrule
\end{tabular}

\end{table}

\begin{table}[H] 
\caption{Average precision and standard deviation values of class 4. The highest average precision with relative lowest standard deviation are marked in \textbf{bold text}.}
\label{tab:wtb4}
\footnotesize
\centering
\begin{tabular}{p{0.45in}p{0.5in}p{0.5in}ccccc}
\toprule
Feature & Similarity Metric & Image Size & $\mathrm{AP}@1$ & $\mathrm{AP}@5$ & $\mathrm{AP}@10$ & $\mathrm{AP}@15$ & $\mathrm{AP}@20$ \\\midrule
DefChars & Cosine & Raw & \textbf{0.71 $\pm$   0.46} & 0.43 $\pm$ 0.38 & 0.35 $\pm$ 0.28 & \textbf{0.30 $\pm$ 0.20} & 0.25 $\pm$ 0.16 \\
DefChars & Euclidean & Raw & \textbf{0.71 $\pm$ 0.46} & \textbf{0.43 $\pm$ 0.34} & \textbf{0.35 $\pm$ 0.27} & 0.29 $\pm$ 0.19 & 0.24 $\pm$ 0.15 \\
DefChars & Jaccard & Raw & 0.21 $\pm$ 0.41 & 0.23 $\pm$ 0.22 & 0.17 $\pm$ 0.13 & 0.17 $\pm$ 0.11 & 0.17 $\pm$ 0.08 \\
DefChars & Manhattan & Raw & \textbf{0.71 $\pm$ 0.46} & 0.42 $\pm$ 0.35 & 0.34 $\pm$ 0.24 & 0.29 $\pm$ 0.18 & \textbf{0.26 $\pm$ 0.14} \\\midrule
Image & MSE & 8 & \textbf{0.25 $\pm$ 0.44} & \textbf{0.08 $\pm$ 0.13} & \textbf{0.05 $\pm$ 0.07} & \textbf{0.05 $\pm$ 0.05} & \textbf{0.04 $\pm$ 0.04} \\
Image & MSE & 20 & 0.21 $\pm$ 0.41 & 0.06 $\pm$ 0.09 & \textbf{0.05 $\pm$ 0.07} & 0.04 $\pm$ 0.05 & \textbf{0.04 $\pm$ 0.04} \\
Image & MSE & 50 & 0.21 $\pm$ 0.41 & 0.05 $\pm$ 0.09 & 0.03 $\pm$ 0.06 & 0.04 $\pm$ 0.05 & \textbf{0.04 $\pm$ 0.04} \\
\bottomrule
\end{tabular}

\end{table}

\end{appendices}
\end{document}